\documentclass[prd,superscriptsize,reprint]{revtex4-1}
\usepackage[utf8]{inputenc} 
\usepackage{graphicx,xcolor,overpic,mathtools}
\usepackage{amsthm,amsmath,amssymb}
\usepackage[colorlinks]{hyperref}
\definecolor{coolblack}{rgb}{0.0, 0.18, 0.39}
\hypersetup{
    colorlinks = false,
   }
\PassOptionsToPackage{normalem}{ulem}
\usepackage{ulem} 
\usepackage{sidenotes}
\usepackage{bm}
\usepackage{bbold}
\usepackage{url}
\usepackage{physics}
\usepackage[makeroom]{cancel}
\usepackage{cleveref}
\usepackage{tensor}
\usepackage{mathrsfs}
\usepackage{slashed}
\usepackage[mode=buildnew]{standalone}
\newcommand{\lag}{\mathcal{L}}

\usepackage{diagbox}

\NewDocumentCommand{\evat}{sO{\bigg}mm}{%
  \IfBooleanTF{#1}
   {\mleft. #3 \mright|_{#4}}
   {#3#2|_{#4}}%
}
\usepackage{enumerate}
\usepackage{marvosym}
\definecolor{azure}{rgb}{0.0, 0.5, 1.0}

\newcommand{\comment}[1]{}

\begin{document}

\title[]{Quantum Skyrmion crystals and the symmetry energy of dense matter }
\author{Christoph Adam}
\author{Alberto Garc\'ia Mart\'in-Caro}
\author{Miguel Huidobro}%
\author{Ricardo V\'azquez}
\affiliation{%
Departamento de F\'isica de Part\'iculas, Universidad de Santiago de Compostela and Instituto
Galego de F\'isica de Altas Enerxias (IGFAE) E-15782 Santiago de Compostela, Spain
}%
\author{Andrzej Wereszczynski}
\affiliation{
Institute of Physics, Jagiellonian University, Lojasiewicza 11, Krak\'ow, Poland
}%

\date[ Date: ]{\today}
\begin{abstract}
The canonical quantization method for collective coordinates in  crystalline configurations of the generalized Skyrme model is applied in order to find the quantum ground state of Skyrmion crystals and study the quantum corrections to the binding energy resulting from the isospin degrees of freedom. This leads to a consistent description of asymmetric nuclear matter within the Skyrme framework and allows us to compute the symmetry energy of the Skyrmionic crystal as a function of the baryon density, and to compare with recent observational constraints. 

\end{abstract}
\maketitle
\tableofcontents

\section{Introduction}
Although the strong interactions are, in principle, completely described by Quantum Chromodynamics (QCD), a derivation of the properties of nucleons, atomic nuclei and nuclear matter from this fundamental theory is still one of the biggest challenges of theoretical physics. This is related to the nonperturbative nature of the low energy regime where the standard expansion in terms of Feynman diagrams loses its applicability. As a consequence, other approaches like lattice QCD or effective phenomenological models, usually tailored for certain regimes, must be applied. 

The Skyrme model \cite{skyrme1962unified} provides a very attractive field theoretical framework for this purpose. On the one hand, it is very well anchored in the underlying fundamental quantum theory \cite{witten1979current} and, therefore, provides a rare opportunity to study the non-perturbative features of strongly interacting matter at all scales, from single baryons and nuclei to neutron stars. Importantly, the model contains a rather limited number of degrees of freedom which, in the simplest version, are simply pions disguised into an $SU(2)$ valued matrix field $U$, and a small number of terms in the action, which translates into an equally small number of free parameters. Then nucleons and atomic nuclei emerge as collective excitations in such a mesonic fluid. Mathematically, they are topological solitons called Skyrmions, whose topological degree can be rigorously identified with the baryon number \cite{Balachandran:1982dw}. Moreover, the relevant quantum numbers are introduced via an appropriate quantization procedure of the corresponding collective coordinates. 

In the last years, there has been significant progress in the application of the Skyrme model to the description of atomic nuclei. The first source of this development is the use of the vibrational quantization. Here the Hilbert state is built not only on the zero modes, as in the standard rigid rotor quantization \cite{adkins1983static}, but also the lightest massive deformations are take into account. This approach elevated the Skyrme model to a quantitative tool for the understanding of excitation bands of light nuclei \cite{battye2009light,Lau:2014baa,Halcrow:2016spb}. In addition, it has very recently been shown how the spin-orbit interaction leading to a phenomenologically consistent nucleon-nucleon force emerges in the Skyrme model \cite{Halcrow:2020gbm}. Secondly, it is now well understood how to reduce the typically quite large binding energies of Skyrmions, which translate into unphysically large binding energies of atomic nuclei. This requires the addition of new, physically motivated terms, e.g., the so-called sextic term \cite{Adam_2010}, generalized potentials \cite{Gillard:2015eia,Gudnason:2016mms}  or new degrees of freedom, e.g., vector mesons \cite{Naya:2018kyi}. 

In a parallel development, the Skyrme model coupled to gravity has been applied to the investigation of neutron stars and their properties. Indeed, in the last few years, astrophysical observations \cite{Miller:2021qha,Riley:2021pdl} and the analysis of gravitational wave measurements \cite{Abbott_2017,Abbott:2020uma,Abbott:2020khf} have led to a big improvement in our understanding of the nuclear matter equation of state (EoS), especially above the saturation density. From the Skyrme model perspective, again the role of the sextic term is crucial, as this part of the action governs the high pressure and density regime \cite{Adam_2015a}. Indeed, it makes Skyrmionic matter much stiffer at extreme conditions, which results in physically acceptable values of the maximal masses of neutron stars. This has been first found in the case where the EoS was motivated by certain limits of the Skyrme model \cite{Adam:2020yfv}. Later on, it has been confirmed in the full Skyrme model computation \cite{Adam:2021gbm}. Remarkably, some Skyrme model-based EoS lead to observables which pass the current available observational data, concerning, e.g., mass-radius curves, tidal deformabilities and quasi-universal relations between the moment of inertia, Love numbers and the quadrupole moment of slowly rotating stars \cite{Adam:2020aza}. 

The starting point in these computations consists in the numerical derivation of the lowest energy periodic Skyrmion solution representing infinite nuclear matter, at a given density, using a variational approach \cite{Adam:2021gbm}. Next, one can find the corresponding EoS by studying how the energy of such solutions depends on the density. Once the EoS is known, the standard Tolman-Oppenheimer-Volkoff (TOV) approach \cite{tolman1939static}, \cite{oppenheimer1939massive} allows us to study the resulting NS. However, all these results have been obtained only in the purely classical limit, where no quantum corrections are taken into account and, importantly, the proton and the neutron are described by the same classical solution, which preserves isospin symmetry. Hence, classical skyrmion crystals would correspond to the so-called symmetric nuclear matter.
However, realistic models of nuclear matter inside neutron stars do not predict an isospin-symmetric state at all. Instead, only a small fraction of protons over total nucleons is generally allowed. 
Indeed, for a nuclear system with total number of baryons $A=N+Z$, where $N$ is the number of neutrons and $Z$ the number of protons, and defining the isospin asymmetry parameter as $\delta=(N-Z)/A=(1-2\gamma)$, with $\gamma$ the proton fraction or ratio between $Z$ and $A$, the binding energy is usually defined as a function of both the baryon density and the asymmetry parameter,
\begin{equation}
    \frac{E}{A}(n_B,\delta)=E_N(n_B)+S_N(n_B)\delta^2+\order{\delta^3},
\end{equation}
where the term $E_N(n_B)$ would be the binding energy of isospin-symmetric matter, and $S_N(n_B)$ is the so-called \emph{symmetry energy}. The symmetry energy is thus a measure of the change in the binding energy of the system as the neutron-to-proton ratio is changed at a fixed value of the total baryon number, and its knowledge is essential to determine the composition of nuclear matter at high densities. However, its dependence on the density has proven difficult to measure experimentally, and usually it is parametrized as an expansion in powers of the baryon density around nuclear saturation $n_0$,
\begin{equation}
    S_N(n_B)=S_0 +\frac{1}{3}L \epsilon +\frac{1}{18}K_{\rm sym}\epsilon^2 +\cdots
\end{equation}
with $\epsilon=(n-n_0)/n_0$, and 
\begin{equation}
    L=3 n_0\pdv{S_N}{n}\eval_{n=n_0}, \quad  K_{\rm sym}=9n_0^2\pdv[2]{S_N}{n}\eval_{n=n_0}
    \label{eq_symetobs}
\end{equation}
the slope and curvature of the symmetry energy at saturation, respectively. The symmetry energy at saturation is well constrained ($S_0\sim 30$ MeV) by nuclear experiments \cite{FiorellaFantina:2018dga}, but the values of the slope
 and higher order coefficients are still very uncertain. However, recent efforts on the analysis of up to date combined astrophysical and nuclear observations have allowed to constrain the value of these quantities with reasonable uncertainty above nuclear saturation \cite{Landry:2021ezp,Tang:2021snt,deTovar:2021sjo,Gil:2021ols,Li:2021thg}.

The main aim of the current work is to carry the description of Skyrmionic matter beyond the classical realm by semiclassically quantizing the isospin DoF of the Skyrme crystal. This allows us to calculate the isospin moment of inertia (hence the symmetry energy) of a given classical solution, taking into account the electric charge neutrality of the crystal, for which the introduction of a neutralising charged lepton background is crucial. The effect of this background is to introduce a certain isospin asymmetry due to $\beta$ equilibrium with the baryonic matter. Finally, we find the EoS of the full Skyrmion crystal plus lepton background system, which models $npe\mu$ matter, and, in particular, the particle fraction of each component as a function of density.

\section{Generalized Skyrme model and classical crystals}
\subsection{The model}
The generalized Skyrme model is defined by the following Lagrangian density
\begin{align}
    \notag \lag = -\frac{f^2_{\pi}}{16}&\Tr \{L_{\mu}L^{\mu} \}  + \frac{1}{32e^2}\Tr \{\left[L_{\mu},L_{\nu}\right]^2 \}  \\[2mm] 
    &- \lambda^2 \pi^4\mathcal{B}_{\mu}\mathcal{B}^{\mu} + \frac{m^2_{\pi} f^2_{\pi}}{8}\Tr \left(U - I \right),
    \label{Lag}
\end{align}
where $L_\mu=U^\dagger \partial_\mu U$ is the left invariant Maurer-Cartan current and the matrix field decomposes as
\begin{equation}
U = \sigma + i \pi_k \tau_k.
 \label{Ufield}
\end{equation}
Here, $\pi_k$ ($k$ = 1, 2, 3) are the pions and $\tau_k$ are the Pauli matrices. The unitarity of the matrix field puts a condition on the fields. Namely, $\sigma^2 + \pi_i\pi_i = 1$. 
Furthermore, $\mathcal{B}^\mu$ is the conserved topological current which, in the standard manner, defines the topological index of maps $U$, i.e., the baryon charge $B$
\begin{equation}
    B = \int d^3x \mathcal{B}^0, \hspace{2mm} \mathcal{B}^{\mu} = \frac{1}{24\pi^2} \epsilon^{\mu\nu\alpha\beta}\Tr\left\{ L_{\nu}L_{\alpha}L_{\beta} \right\}.
    \label{TopoNumber}
\end{equation}
It is a remarkable property of the generalized Skyrme effective model that it contains only four terms, which translate into four coupling constants $f_\pi, m_\pi, e, \lambda$.  Two of them have a direct phenomenological interpretation as the pion decay constant and the pion mass. In addition, $\lambda$ can be related to a ratio between the mass and the coupling constant of the $\omega$ meson. We assume the physical mass of the pions, $m_\pi=140$ MeV, in the whole paper. The parameters of the standard Skyrme model are set to
\begin{equation} \label{standard-values}
f_{\pi} = 129 \:\text{MeV}\, , \quad e = 5.45. 
\end{equation}
These values of $f_\pi $ and $e$ are often used in the Skyrme model literature \cite{castillejo1989dense,Baskerville:1996he}, because they were obtained in \cite{adkins1983static} by fitting the $B = 1$ solution of the standard Skyrme model to the masses of the proton and the $\Delta$ excitation. Our main reason to choose them is that this choice facilitates the comparison of our calculations with previous results. Finally, we will consider different values for the coupling constant $\lambda^2$ multiplying the third term in the action, containing six powers of derivatives. Although often omitted in the context of light nuclei, this term is essential when one studies the properties of nuclear matter at high density, which is a natural environment in the core of neutron stars. Indeed, this sextic term governs the equation of state at this regime and asymptotically leads to the maximally stiff EoS \cite{Adam:2015lra}. 

\subsection{The classical crystal of Skyrmions} \label{Sec: Crystal ansatz}
In the Skyrme model, infinite skyrmionic matter is described by a periodic minimizer of the static energy
\begin{align}
    \notag E = \frac{1}{24\pi^2}&\int d^3x \left[ -\frac{1}{2}\Tr \{L_iL_i\} - \frac{1}{4}\Tr \{\left[L_i,L_j\right]^2\}  \right.\\[2mm]
    &\left. + 8\lambda^2 \pi^4 \frac{f^2_{\pi} e^4}{\hbar^3} (\mathcal{B}^0)^2 + \frac{m^2_{\pi}}{f^2_{\pi}e^2}\Tr\left( I - U \right)  \right] \label{Energy} 
\end{align}
and, therefore, it is usually referred to as the Skyrme crystal. For numerical purposes, we have adopted Skyrme units, in which energy and length are measured in units of $3\pi^2 f_{\pi}/e$ and $\hbar/(f_{\pi}e)$, respectively.
Obviously, while the total energy of the crystal is infinite, the energy per baryon number remains finite
\begin{equation}
    \frac{E}{B} = \frac{N_{\text{cells}}\:E_{\text{cell}}}{N_{\text{cells}}\:B_{\text{cell}}}=\frac{E_{\text{cell}}}{B_{\text{cell}}}.
\end{equation}
Here, $N_\text{cells}$ is the number of cells and $E_\text{cell}$, $B_\text{cell}$ are the energy and baryon charge in a single, periodic cell. The energy of the unit cell strongly depends on the assumed geometry and its size, characterized by the length parameter $L$. Concretely, $L$ is the distance between nearest-neighbor skyrmions in the maximally attractive channel. As a consequence, the resulting field configuration is {\em not} periodic in $L$. The period length and the size of the unit cell is $2L$, instead. 

Comparing various geometries, i.e., types of crystals, at a particular volume of the cell, one can find the ground state crystalline solution at a given density. Note that a particular geometry of the crystal translates into particular symmetries of the Skyrme field. In practice, it is impossible to check all possible geometries and one constrains considerations to well motivated cubic crystals. These are the simple cubic (SC) and face centered (FCC) crystals of Skyrmions as well as the body centered (BCC) and face centered (FCC) crystals of {\it half-Skyrmions}. 

As the cubic symmetries are shared by all these geometries, the following relations must be obeyed by these crystals,
\begin{align}
    \notag\text{A}_1&: (x,y,z) \rightarrow (-x,y,z), \\[2mm] &(\sigma,\pi_1,\pi_2,\pi_3) \rightarrow (\sigma,-\pi_1,\pi_2,\pi_3), \label{A_1}\\[2mm]
    \notag\text{A}_2&: (x,y,z) \rightarrow (y,z,x), \\[2mm] &(\sigma,\pi_1,\pi_2,\pi_3) \rightarrow (\sigma,\pi_2,\pi_3,\pi_1).
    \label{A_2}
\end{align}

In this work we will focus on the FCC crystal of half-skyrmions, which has two additional symmetries,
\begin{align}
    \notag\text{C}_3&: (x,y,z) \rightarrow (x,z,-y), \\[2mm] &(\sigma,\pi_1,\pi_2,\pi_3) \rightarrow (\sigma,-\pi_1,\pi_3,-\pi_2), \\[2mm]
    \text{D}_4&: (x,y,z) \rightarrow (x+L,y,z), \\[2mm] &(\sigma,\pi_1,\pi_2,\pi_3) \rightarrow (-\sigma,-\pi_1,\pi_2,\pi_3).
\end{align}

A more detailed description of the construction of the Skyrme crystal and the comparison of different symmetries can be found in \cite{Adam:2021gbm}, and we have kept the same notation for this work. As in that previous work, the unit cell has size $2L$ and a baryon content of $B_{\text{cell}} = 4$. Then, for each value of $L$ we obtain the minimum of energy as explained in \cite{Adam:2021gbm}. It turns out that the energy-size curve, $E_{\text{cell}}(L)$, is a convex function which has a minimum at a certain $L_*$. 

\subsection{The Skyrme crystal EoS for symmetric nuclear matter}

For the crystal solutions one can define the relevant thermodynamical quantities in the usual way, that is, energy density $\rho$, pressure $p$ and baryon charge density $n_B$
\begin{align}
    \rho &= \frac{E}{V} = \frac{E_{\text{cell}}}{V_{\text{cell}}}, \\[2mm]
    p &= - \frac{\partial E}{\partial V}= -\frac{\partial E_{\text{cell}}}{\partial V_{\text{cell}}} , \\[2mm]
    n_B &= \frac{B}{V}= \frac{B_{\text{cell}}}{V_{\text{cell}}}.
\end{align}
Again they are functions of $L$ or, in other words, the volume of the unit cell $V_{\text{cell}}$. 

At the point where $L=L_*$, the given crystal solution describes skyrmionic matter at equilibrium, i.e., at zero pressure. In the region where $L < L_*$ the volume of the cell decreases, which corresponds to a squeezed crystal. This translates into growing pressure and density. The remaining region $L > L_*$, where the volume increases in comparison to the equilibrium, is thermodynamically unstable. Indeed, it formally gives negative pressure. Due to that, the low density regime cannot be described by any of the previously mentioned crystals. On the contrary, it is expected that the crystal is replaced by an inhomogeneous phase, where lumps of nontrivial energy density are surrounded by regions of void. 

In the absence of the sextic term, the landscape of the crystal energy minimizers was very well understood. Namely, for $L\leq L_*$ the ground state is formed by the FCC half-Skyrmion phase, which at extremely high density is replaced by the BCC half-Skyrmions. This phase transition occurs at densities much beyond the values expected at the cores of neutron stars. 

It has been recently shown \cite{Adam:2021gbm} that this picture is significantly modified if the sextic term is added. First of all, the FCC to BCC phase transition is moved to much smaller densities, approximately 4-5 saturation densities, which can be easily found in the center of heavy NS. In addition, the appearance of the sextic term introduces a fluidity into the model which mathematically results from the volume diffeomorphism invariance of this part of the action. Such a fluidity is visible in a more and more homogenous distribution of the energy density as the pressure increases. This does not happen if this term is absent. 

However, perhaps the most crucial result is that the equation of state, which relates energy density and pressure, stiffens in the generalized Skyrme model. Indeed, the sextic term alone leads to the maximally stiff EoS, $\rho = p$. In the full, generalized model it occurs asymptotically at high density. In any case, this stiffening is responsible for a significant rising of the maximal masses of NS to values which are in accordance with current observations. 

\comment{
The resulting EoS for the generalized Skyrme model is shown in Fig. \ref{Figure.EoS} {\color{red} where also the FCC to BCC phase transition is very well visible}.

\begin{figure}[h!]
    \centering
    \includegraphics[scale=0.36]{Figures/EoS.pdf}
    \caption{The EoS $\rho(p)$ and $n(p)$ for the symmetric nuclear matter in the generalized Skyrme model. The FCC to BCC phase transition occurs at $p=...$.}
    \label{Figure.EoS}
\end{figure}
}
These very encouraging findings have been obtained in the classical regime of the Skyrme model, which, as we already underlined, corresponds to symmetric nuclear matter. Therefore, it is of vital importance to semiclassically quantize the isospin DoF of the skyrmionic crystal. This is the first step to describe neutron matter, which is the ground state of the matter in the core of NS. 

\section{Quantization of the Skyrme crystal}
\label{sec:Quantization}
It is well known that the largest quantum correction to the classical energy of Skyrmion configurations comes from the contribution of the isospin degrees of freedom, which are usually quantized as zero-modes via some collective coordinate parametrization. To add the contribution of the quantization of the global isospin zero modes to the total energy, we need to know the quantum isospin state of the full crystal. This task, however, becomes impossible in the thermodynamic limit in which the number of particles forming the crystal goes to infinity. Instead, we can make the following assumptions on the quantum wavefuntion of the full crystal:
\begin{itemize}
    \item The isospin wavefuntion of the total crystal $\ket{\Psi}$ can be written as a superposition of states constructed from the (infinite) product of isospin wavefuntions of individual unit cells, $\ket{\Psi}=\bigotimes\limits_{\rm cells}\ket{\psi}$. In other words, as a first approximation we will not consider the quantum correlation on the isospin state between cells. 
    \item The symmetry of the classical crystal configuration is inherited by the total wavefunction, and shared with the wavefunction of each of the unit cells, i.e. both $\ket{\Psi}$ and $\ket{\psi}$ share the same point symmetry group.
\end{itemize}
These two assumptions imply that finding the quantum states of the total crystal is equivalent to finding the state of each unit cell. The latter is in fact a more plausible task as we may use the tools developed for the quantization of multi-skyrmion configurations.

\subsection{The internal symmetry of Skyrme crystals and the isospin group}
The full symmetry group of the massless Skyrme Lagrangian is given by the direct product of the Poincar\'e and chiral groups. However, we are interested in solutions that minimize the energy functional (measured on a given reference frame). The internal symmetry group of such functional is the same as the Lagrangian, but the Poincar\'e symmetry is broken to the Euclidean subgroup corresponding to spatial rotations and translations, $E_3=SO(3)\times \mathbb{R}^3$. Thus, the symmetry group of the energy functional is $\widetilde{G}=E_3\times SU(2)_L\times SU(2)_R\simeq E_3\times SO(4)_{chiral}$. The action of an element of such group on the Skyrme field is given by
\begin{equation}
    U(\Vec{x})\rightarrow g_L U(R_S\cdot \Vec{x}+\vec{a})g_R^\dagger
    \label{group_action}
\end{equation}
where $\vec{a}\in \mathbb{R}^3$, $R_S\in SO(3)$ represents the spatial rotations and $g_{L/R}\in SU(2)_{L/R}$ are the left and right-handed chiral transformations, respectively. 

Moreover, the presence of nontrivial boundary conditions imposed on the relevant field configurations may further reduce their symmetry. Indeed, in the case of finite energy field configurations a boundary condition of the form $U(\Vec{x})\xrightarrow{|\vec{x}|\rightarrow \infty} \mathbb{1}$ must be imposed (i.e. the Skyrme field must decay to its vacuum value far from the center of the soliton). Such boundary condition not only allows us to classify the field configurations into different topological sectors labeled by their homotopy class within the third homotopy group of the target space $\pi_3(SU(2))=\mathbb{Z}$, but also reduces the symmetry of such configurations since the vacuum is only preserved under the subgroup $G=E_3\times \rm{diag}[SU(2)_L\times SU(2)_R]\simeq E_3\times SU(2)_I$, i.e. transformations of the form \eqref{group_action} with $g_L=g_R=g\in SU(2)_I$. The remaining internal symmetry group is called the \emph{isospin group}, since it corresponds to the isospin degrees of freedom when the solitons (Skyrmions) are identified with baryons and nucleons of low-energy QCD and nuclear physics.

A further reduction of the symmetry group $G$ may occur on individual configurations minimizing the energy functional for each topological sector. For instance, the $B=1$ Skyrmion does present the full group $G$ as a symmetry of its energy density isosurfaces, whereas the spatial rotations are broken to $O(2)=SO(2)\times \mathbb{Z}_2$ for the $B=2$ Skyrmion, which presents a toroidal shape. As $B$ increases, the symmetry of the configurations minimizing the static energy becomes more complicated, and for $B\geq 3$ it is given by a point group, the $B=3$ Skyrmion presenting tetrahedral symmetry, the $B=4$ cubic symmetry and so on.

 On the other hand, for crystalline configurations, the boundary conditions imposed on the Skyrme field are of different nature: instead of imposing the vacuum at large distances, one must impose periodicity conditions on the boundary of a compact space region- the \emph{unit cell}- and hence these solutions do not yield a finite value when its energy density is integrated over all space. Instead, this requirement is relaxed to yield a finite value of the energy over the unit cell, or, equivalently, a finite energy per baryon.
 
 Therefore, the symmetry group of a crystalline configuration is reduced from $\widetilde{G}$ due to these periodicity conditions, which in turn result in a particular point group symmetry for the full crystal. Indeed, in section \ref{Sec: Crystal ansatz} we have studied different crystalline structures that present different symmetries, although all of them are based on a simple cubic symmetry. However, it is not straightforward to define the isospin subgroup $SU(2)_I$ in such configurations, since the vacuum value is no longer imposed at the boundaries. Instead, one should in principle consider internal rotations of the full chiral group $SO(4)_{\rm chiral}$, since there is not a natural way to select the diagonal subgroup corresponding to isospin.
 
 The problem of defining the isospin group in crystal configurations is treated in \cite{Baskerville:1996he}. As explained there, the procedure of defining the isospin subgroup in crystalline configurations is subject to some ambiguities, but the energy per baryon spectrum in the infinite crystal limit is unique. Indeed, it turns out that the 4-dimensional representation of the cubic point group of minimal energy crystals is \emph{reducible} into the trivial 1-D irrep and a 3-D irrep, which singles out one direction in isospin space. We may then choose the $\sigma$ field to transform in the trivial irrep, and then to define the isospin group as the subgroup of isorotations within the 3-D irrep, i.e. rotations between the three pion fields.
 
 In the rest of this work, we will use this definition of isospin in crystals, as it is also the natural choice if a symmetry-breaking potential (such a mass term for the pions) is added to the Skyrme Lagrangian.

\subsection{Quantum isospin states and Hilbert space}
In general, the classical field configuration of a $B>1$ Skyrmion presents a non-trivial symmetry
characterized by a finite point group $H$. Hence, the corresponding allowed quantum states of spin and isospin will correspond to different linear combinations of the basis vectors $\ket{ii_3k_3}$ (see \cref{append_matrixelements}).
To find these combinations explicitly, the group $H$ has to be known in terms of a set of generators, $\{H_k\}$ that can be written as a product of a rotation $\mathcal{R}_k$ and an isorotation $\mathcal{R'}_k$. Then, the Finkelstein-Rubinstein (F-R) constraints may be written as
\begin{equation}
    \exp{i\alpha \mathbf{n}\mathbf{L}}\exp{i\beta \mathbf{N}\mathbf{K}}\ket{\psi} = \chi_{FR}\ket{\psi},
    \label{FRconstraints}
\end{equation}
where $\mathbf{L}$ and $\mathbf{K}$ are the (body-fixed) angular and isospin angular momentum operators, respectively, and $\chi_{FR} = \pm 1$. 
To consistently do this, it is important to know as well whether each of the symmetries $H_k$ corresponds to a contractible or a non-contractible loop in configuration space. Once all the generators $H_k$ have been parametrized in terms of products of rotations and isorotations, one needs to find a solution of the Finkelstein-Rubinstein system of constraints on each subspace of fixed $i,j$. Solving the system of F-R constraints is equivalent to finding a set of common eigenvectors in this subspace  of a particular set of matrices. 

Indeed, the general transformation law of angular momentum states under a rotation
$\mathcal{R}(\alpha,\beta,\gamma)$  parametrized in terms of the Euler angles $\alpha,\beta,\gamma$ (in the $Z-Y-Z$ convention) is
\begin{equation}
    \mathcal{R}(\alpha,\beta,\gamma)\ket{j,m}=\sum\limits_{m'}D^j_{mm'}(\alpha,\beta,\gamma)\ket{jm'},
    \label{defrots}
\end{equation}
where $D^j_{mm'}(\alpha,\beta,\gamma)$ is the Wigner D-matrix, corresponding to the irreducible, spin-$j$ representation of the rotation group.

Hence, for each $H_k$, a solution of the corresponding F-R constraint is given by a state $\ket{\Psi}$ which is both an eigenstate of $\mathcal{R}_k(\alpha,\beta,\gamma)$ with eigenvalue $\lambda_k$ and of $\mathcal{R}'_k(\alpha',\beta',\gamma')$ with eigenvalue $\lambda'_k$, and such that $\lambda_k\times \lambda'_k=\chi_{\rm{FR}}$. On each subspace given by a fixed value of total spin and isospin, this is equivalent to finding a common eigenvector of the corresponding Wigner D-matrices. The allowed (physical) states will be those which satisfy all the F-R constraints.

In the case of a Skyrmion crystal, the unit cell presents a concrete set of symmetries, some of which relate rotations both in space and isospace, and hence we should consider as physical states only those that are compatible with such symmetries. Let us now proceed to calculate the Finkelstein-Rubinstein (FR) constraints in order to obtain the corresponding quantum states of the Skyrme crystal. The relevant symmetries of the FCC half-skyrmion crystal linking rotations and isorotations are $A_2$ and $C_3$, which are represented by the following operators,
\begin{align}
    &\exp{i \frac{\pi}{2}\frac{1}{\sqrt{3}}\left( K_1 + K_2 + K_3 \right)} = \mathcal{R}'\left( 0, -\pi/2, -\pi/2 \right), \\[2mm]
    &\exp{i \frac{\pi}{2}K_1} = \mathcal{R}'\left( \pi/2, -\pi/2, \pi/2 \right).
    \label{symmetriesFR}
\end{align}
We only write the operators that correspond to isospin transformations, since the rotations in the real space are the same. Recall that we are using the Euler angles representation for the rotation and isorotation operators in the $ZYZ$ convention,
\begin{equation}
    \mathcal{R}\left(\alpha, \beta, \gamma \right) = R_z\left( \alpha \right)R_y\left( \beta \right)R_z\left( \gamma \right).
\end{equation}
From \eqref{defrots} we know how these operators act on a state $\ket{j, l_3} \otimes \ket{i, k_3}$ (again we will only consider the isospin part, the quantum numbers resulting from the spin quantization are the same),
\begin{equation}
    \mathcal{R}\left(\alpha, \beta, \gamma \right) \ket{i, k_3} = \sum_{k'_3} D^i_{k_3, k'_3}\left(\alpha, \beta, \gamma \right)\ket{i, k'_3},
\end{equation}
where $D^I_{k_3, k'_3}$ are the Wigner D-matrices. Then we can consider this as a problem of finding the eigenvalues and eigenvectors of the Wigner D-matrices, and the quantum states will be the combination of $J, L_3, I, K_3$ that satisfy the FR constraints \eqref{FRconstraints}. In our case, we will consider the possible quantum states of a unit cell, which carries a baryon number $B_{\rm cell} = 4$. We will not prove here that the transformations \eqref{FRconstraints} with the corresponding spatial rotations correspond to contractible loops on the configuration space. Instead, we will assume that this is indeed the case, i.e. $\chi_{FR} = +1$, as tends to be the case for $B=0 \mod 4$ Skyrmions \cite{Krusch:2005iq}.  Also, we will consider all the possible values of $i$, namely $i=0,1,2$ (eigenvalue of the isospin moment of inertia) and show the (unnormalized) eigenvectors for each symmetry.
\\
\\

\paragraph{$A_2$ symmetry: $\mathcal{R}(0,-\pi/2,-\pi/2)$}
\begin{itemize}
    \item For $i = 0$, $D^0_{00} = 1$ and
    there is only one state $\ket{i = 0,i_3 = 0}$.
    
    \item For $i = 1$, we have
    \begin{equation}
        D^1_{k_3, k'_3} = \begin{pmatrix}
        i/2 & -1/\sqrt{2} & -i/2\\
        i/\sqrt{2} & 0 & i/\sqrt{2}\\
        i/2 & 1/\sqrt{2} & -i/2
        \end{pmatrix}.
    \end{equation}
This matrix has three different eigenvalues and corresponding eigenstates
    \begin{align}
        &\lambda_1 = 1,\notag\\
         \qquad&\ket{\psi^1_1} = -i\ket{1,-1} + (1 + i)/\sqrt{2}\ket{1,0} + \ket{1,1}\notag\\
        &\lambda_2 = -1/2 - \sqrt{3}/2i,\notag\\
         \qquad&\ket{\psi^1_2} = (2-\sqrt{3})i\ket{1,-1} - \sqrt{2-\sqrt{3}}(1 + i)\ket{1,0} + \ket{1,1}\notag\\
        &\lambda_3 = -1/2 + \sqrt{3}/2i,\notag\\ \qquad&\ket{\psi^1_3} = (2+\sqrt{3})i\ket{1,-1} + \sqrt{2+\sqrt{3}}(1 + i)\ket{1,0} + \ket{1,1}
    \end{align}
    
    \item For $i = 2$, we have
    \begin{equation}
        D^2_{k_3, k'_3} = \begin{pmatrix}
        -1/4 & -1/2i & \sqrt{6}/4 & 1/2i & -1/4\\
        -1/2 & -1/2i & 0 & -1/2i & 1/2\\
        -\sqrt{6}/4 & 0 & -1/2 & 0 & -\sqrt{6}/4\\
        -1/2 & 1/2i & 0 & 1/2i & 1/2\\
        -1/4 & 1/2i & \sqrt{6}/4 & -1/2i & -1/4
        \end{pmatrix}
    \end{equation}
    which shares the same eigenvalues as the corresponding $i=1$ case, but this time $\lambda_2$ and $\lambda_3$ present multiplicity two. The corresponding eigenstates are
    \begin{align}
    & \lambda_1 = 1,\notag \\
        &\ket{\psi^2_1} = -\ket{2,-2} + (1-i)\ket{2,-1} + (1+i)\ket{2,1} + \ket{2,2},\\[2mm]
       & \lambda_2 = -1/2 - \sqrt{3}/2i, \notag\\
        &\ket{\psi^2_{2a}} = \ket{2,-2} - \sqrt{2}i\ket{2,0} + \ket{2,2}, \\ &\ket{\psi^2_{2b}} = -\ket{2,-2} - \frac{(1+\sqrt{3})}{2}(1-i)\ket{2,-1} + \notag\\
        &\hspace{1.3cm}+ \frac{(\sqrt{3}-1)}{2}(1+i)\ket{2,1} + \ket{2,2},\notag\\[2mm]
       & \lambda_3 = -1/2 + \sqrt{3}/2i, \notag\\
        &\ket{\psi^2_{3a}} = \ket{2,-2} + \sqrt{2}i\ket{2,0} + \ket{2,2}, \\ &\ket{\psi^2_{3b}} = -\ket{2,-2} + \frac{(\sqrt{3}-1)}{2}(1-i)\ket{2,1} - \notag\\
        &\hspace{1.3cm}-\frac{(1+\sqrt{3})}{2}(1+i)\ket{2,1} + \ket{2,2}.\notag
    \end{align}
\end{itemize}
\paragraph{$C_3$ symmetry: $\mathcal{R}'\left( \pi/2, -\pi/2, \pi/2 \right)$}
\begin{itemize}
    \item Again, for $i = 0$, the only state is $\ket{0,0}$.
    
    \item For $i = 1$, the corresponding Wigner matrix
    \begin{equation}
        D^1_{k_3, k'_3} = \begin{pmatrix}
        -1/2 & i/\sqrt{2} & 1/2\\
        -i/\sqrt{2} & 0 & -i/\sqrt{2}\\
        1/2 & i/\sqrt{2} & -1/2
        \end{pmatrix}
    \end{equation}
    has two eigenvalues, with multiplicity $2$ and $1$, respectively. The associated eigenstates are
    \begin{align}
        &\lambda_1 = -1,\notag \\&\ket{\phi^1_{1a}} = -\ket{1,-1} + \ket{1,1},\\ &\ket{\phi^1_{1b}} = \ket{1,-1} + \sqrt{2}i\ket{1,0} + \ket{1,1}\notag\\[2mm]
        &\lambda_2 = 1, \notag\\&\ket{\phi^1_2} = \ket{1,-1} - \sqrt{2}i\ket{1,0} + \ket{1,1}
    \end{align}
    
    \item $i = 2$,
    \begin{equation}
        D^2_{k_3, k'_3} = \begin{pmatrix}
        1/4 & -i/2 & -\sqrt{6}/4 & i/2 & 1/4\\
        1/2i & 1/2 & 0 & 1/2 & -i/2\\
        -\sqrt{6}/4 & 0 & -1/2 & 0 & -\sqrt{6}/4\\
        -i/2 & 1/2 & 0 & 1/2 & i/2\\
        1/4 & i/2 & -\sqrt{6}/4 & -i/2 & 1/4
        \end{pmatrix}
    \end{equation}
    \begin{align}
        \lambda_1 = -1,& \\
        &\ket{\phi^2_{1a}} = -\ket{2,-2} + i\ket{2,-1} - i\ket{2,1} + \ket{2,2}, \nonumber\\
        &\ket{\phi^2_{1b}} = \ket{2,-2} + \sqrt{6}\ket{2,0} + \ket{2,2}\notag\\
    \lambda_2 = 1,& \\
        \qquad &\ket{\phi^2_{2a}} = \ket{2,-1} + \ket{2,1},\nonumber \\
         \qquad &\ket{\phi^2_{2b}} = -\ket{2,-2} - 2i\ket{2,-1} + \ket{2,2}, \nonumber\\
         \qquad &\ket{\phi^2_{2c}} = \ket{2,-2} - \sqrt{2/3}\ket{2,0} + \ket{2,2}.\nonumber
    \end{align}
\end{itemize}
Physical states with fixed $i$ will  correspond to mutual eigenstates of both spin-$i$ Wigner matrices corresponding to the two symmetries, $A_2$ and $C_3$. For each value of total isospin, we have been able to find a unique state that satisfies this property,
\begin{equation}
    \begin{split}
        i=0&\rightarrow \ket{0,0},\\
        i=1&\rightarrow \ket{\psi^1_1}=(1+i)/2\ket{\phi^1_{1a}}+(1-i)/2\ket{\phi^1_{1b}},\\
        i=2&\rightarrow \ket{\psi^2_1}=\ket{\phi^2_{2b}}+(1+i)\ket{\phi^2_{2a}}
    \end{split}
\label{quantumstates}
\end{equation}
and the corresponding normalized states will be denoted by $\ket{\psi^i}$. Note that each $\ket{\psi^i}$ corresponds to an isospin multiplet with degeneracy $i$, since the $i_3$ quantum number is not restricted by the F-R constraints (and the same happens to the corresponding spin states, in which $j_3$ is not constrained either).

Then, we just find the quantum ground state of the crystal as the state with lower total energy. Naively, one would think that it corresponds to the $\ket{\psi^0}=\ket{0,0,0}$ state, as the isospin contribution vanishes. However, this would result in a non-vanishing total electric charge of each unit cell, so the crystal will be unstable due to an infinite contribution of the Coulomb energy, as was already noticed in \cite{klebanov1985nuclear}. Thus, electrical neutrality implies that the true ground state of pure skyrmion matter corresponds to the $i_3=-2$ state in the multiplet $\ket{\psi^2}$.

Following the previous reasoning we have obtained the allowed quantum states for a single unit cell. On the other hand, we have assumed that a basis for the Hilbert space of the total crystal isospin state can be constructed from the direct product of states of individual cells. We are now in a position to be more specific about this statement. Indeed, consider the quantum state of two unit cells with $i=\{i_1,i_2\}$ and $i_3 =\{ m_1,m_2\}$. The allowed values of $\{i_1,i_2\}$ are $\{0,1,2\}$ , and $-i_a\leq m_a\leq i_a$, $a=1,2$. The total quantum state will be an eigenstate of the total isospin, so it will be better described in the coupled angular momenta basis. Indeed, from representation theory, the tensor product of two spin $j_a$ representations may be decomposed as a direct sum as
\begin{equation}
    D^{j_1}\otimes D^{j_1}=\bigoplus\limits_{k=\abs{j_2-j_1}}^{j_2+j_1} D^k.
\end{equation}
Thus, the basis of states for the two-cell system will be $\ket{I,I_3,i_1,i_2}$, with $\abs{j_1-j_2}\leq I\leq j_1+j_2$ the total isospin number and $-I\leq I_3\leq I$ the third component of total isospin. Therefore, to find a general basis state for an arbitrary number of unit cells, we should just need to generalize the previous construction to the coupling of an arbitrary number of different angular momenta. There are arbitrarily many ways to do it, which should be equivalent up to a unitary transformation (at least, for an arbitrary, but finite, number of unit cells). An important final remark is that, following such construction, the total isospin and third component of isospin of the full crystal will remain good quantum numbers, independently of the quantum state, so they will correspond to well defined observables in the quantum theory, as opposed to the isospin of each of the individual cells.

\subsection{Isospin correction to the energy per baryon}
\label{sec:Isospin}
Let us rewrite the Skyrme Lagrangian \eqref{Lag} as
\begin{align}
    \notag \mathcal{L}=\frac{1}{24\pi^2}&\big[a\Tr\{L_{\mu}L^{\mu}\} + b\Tr\{\left[L_{\mu},L_{\nu}\right]^2\} \\
     &+c\,\mathcal{B}_{\mu}\mathcal{B}^{\mu} + d\Tr (U - I )\big].
\end{align}
The values of $a,b,c,d$ are easily obtained from \eqref{Energy},
\begin{equation}
    a=-\frac{1}{2}, \quad b=\frac{1}{4},\quad c=-8\lambda^2\pi^4\frac{f_\pi^2e^4}{\hbar^3}, \quad 
    d = \frac{m^2_{\pi}}{f^2_{\pi}e^2}.
\end{equation}
We now consider a (time-dependent) isospin transformation of a static Skyrme field configuration,
\begin{equation}
    U(\vec{x})\rightarrow \tilde U(\vec{x},t)\equiv g(t) U(\vec{x}) g^\dagger(t).
    \label{IsorotF}
\end{equation}
The Maurer-Cartan (M-C) form transforms as ($\dot{g}=dg/dt$)
\begin{equation}
    \Tilde{U}^\dagger\partial_\mu \Tilde{U}=\left\{\begin{array}{ll}
         g U^\dagger\partial_i U g^\dagger, &  (\mu=i=1,2,3)\\
         g(U^\dagger[g^\dagger \dot{g},U])g^\dagger,&(\mu=0) .
    \end{array}\right. 
    \label{MCiso}
\end{equation}
We now define the isospin angular velocity $\vec{\omega}$ as $g^\dagger\dot{g}=\tfrac{i}{2}\omega_a\tau_a$. Then, we may write the time component of the Maurer-Cartan current as
$ \tilde U^\dagger\partial_0 \tilde U=g T_ag^\dagger \omega_a$, where $ T_a$ is the $\mathfrak{su}(2)$-valued current,
\begin{equation}
    T_a=\frac{i}{2}U^\dagger[\tau_a,U]=i(\pi_a\pi_b-\pi_c\pi_c\delta_{ab}+\sigma\pi_c\epsilon_{abc})\tau_b\equiv iT_b^a\tau_b,
    \label{UdcommtU}
\end{equation}
where we have made use of the parametrization \eqref{Ufield}. Moreover, the spatial components of the M-C form can be written in terms of the sigma and pion fields as well,
\begin{equation}
\begin{split}
    L_k=&(\sigma-i\pi_a\tau_a)(\partial_k\sigma+i\partial_k\pi_b\tau_b)=\\
    =&i(\sigma\partial_k\pi_c-\pi_c\partial_k\sigma+\pi_a\partial_k\pi_b\varepsilon_{abc})\tau_c\equiv L_k^c\tau_c.
\end{split}
    \label{LeftMC}
\end{equation}

The time dependence of the new Skyrme field induces a kinetic term in the energy functional, given by \footnote{Remember that we are using the mostly minus convention for the metric signature.}
\begin{align}
    \notag T=\frac{1}{24\pi^2}\int\{&a\Tr\{L_0L_0\}-2b\Tr\{[L_0,L_k][L_0,L_k]\} \\
    &-c\,\,\mathcal{B}^i\mathcal{B}_i\}d^3x,
\end{align}
with $\mathcal{B}^i$ the spatial components of the topological current:
\begin{equation}
    \mathcal{B}^i=\frac{1}{24\pi^2}\varepsilon^{i\alpha\beta\gamma}\Tr\{L_\alpha L_\beta L_\gamma\}=\frac{3}{24\pi^2}\varepsilon^{ijk}\Tr\{L_0L_jL_k\}.
\end{equation}
We may rewrite the kinetic isorotational energy in the standard way as a quadratic form acting on the components of the isospin angular velocity,
\begin{equation}
    T=\frac{1}{2}\omega_i\Lambda_{ij}\omega_j
\end{equation}
where $\Lambda_{ij}$ is the isospin inertia tensor, given by
\begin{equation}
\begin{split}
    \Lambda_{ij}=\frac{1}{24\pi^2}&\int d^3x\left\{2a\Tr\{T_iT_j\}-4b\Tr\{[T_i,L_k][T_j,L_k]\}\right.\\
    -&\left.\frac{c}{32\pi^4}\varepsilon^{abc}\Tr\{T_iL_bL_c\}\varepsilon_{ars}\Tr\{T_jL_rL_s\}\right\}\,.
    \label{Inertia_Tensor}
\end{split}
\end{equation}
As shown in \cref{append:inertia-tensor}, the complete isospin inertia tensor for the unit cell of a cubic crystal will be proportional to the identity, and its eigenvalue (the isospin moment of inertia) will be given by 
\begin{equation}
    \Lambda =\frac{1}{24\pi^2} \left[2a \Lambda^{(2)}-4b\Lambda^{(4)}-\frac{c}{32\pi^4}\Lambda^{(6)} \right].
    \label{iso_moment}
\end{equation}

\begin{figure*}[h!]
    \centering
    \includegraphics[scale=0.45]{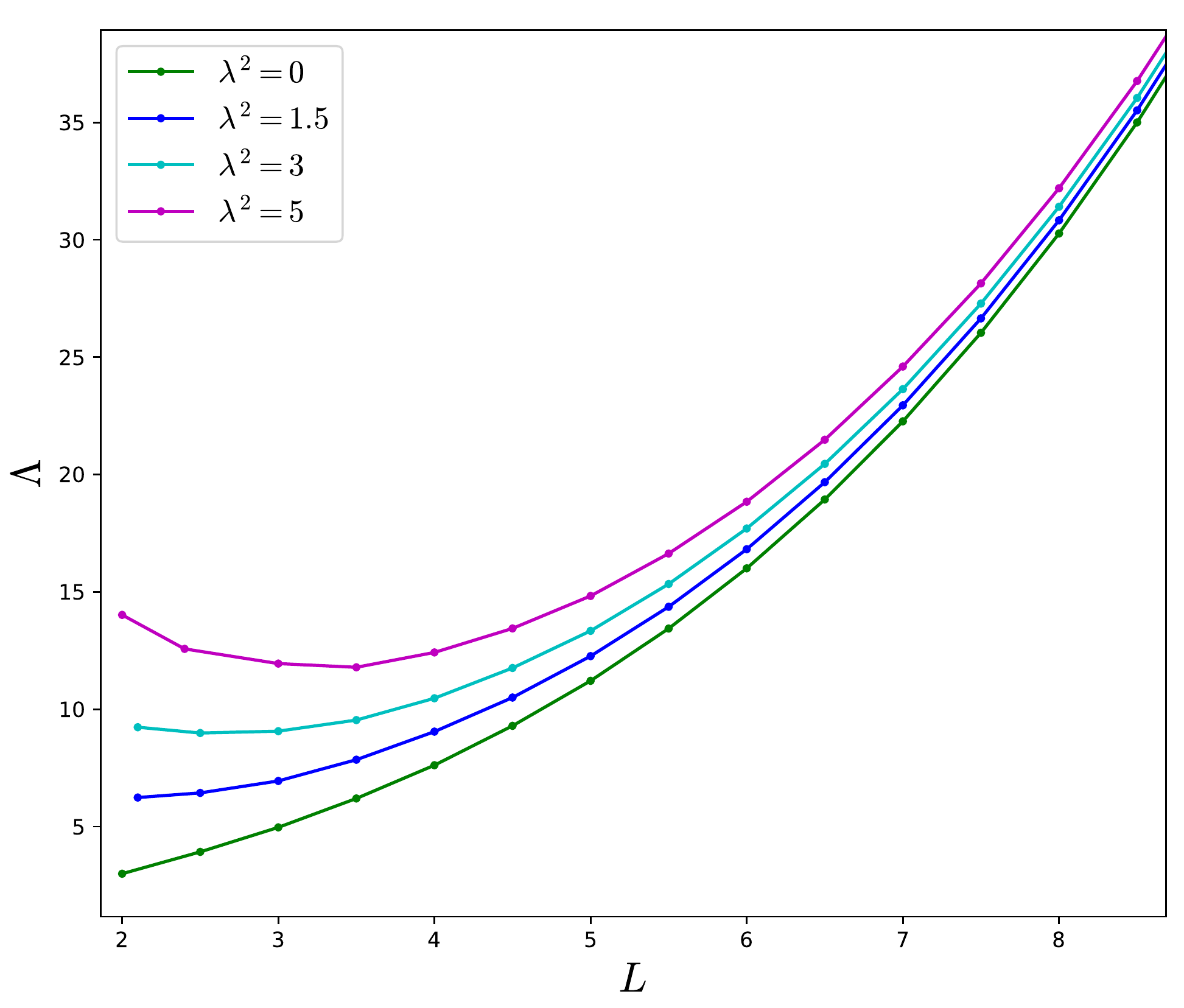}
    \caption{The eigenvalue $\Lambda$ of the isospin inertia tensor is displayed against the lattice length parameter in Skyrme units.}
    \label{fig:LambdaIsospin}
\end{figure*}

The numerical results for $\Lambda$ for the $\mathcal{L}_{240}$, and the full $\mathcal{L}_{2460}$ cases are plotted as a function of the lattice parameter 
in \cref{fig:LambdaIsospin}, for different values of the sextic term coupling constant $\lambda^2$. The $\Lambda$ curve for $\lag_{24}$ was obtained by Baskerville in \cite{Baskerville:1996he} so it has been helpful to check our results. The value of $\Lambda$ becomes smaller at high densities both without sextic or with a small value of $\lambda^2$, hence the isospin correction to the energy will grow with $n_B$. However the increase of the sextic coupling constant produces a final increase of $\Lambda$ at high densities (at which the sextic term becomes more relevant), so a non-trivial behavior will be found in the symmetry energy. Therefore, the result for $\mathcal{L}_{2460}$ diverges from that for  $\mathcal{L}_{240}$ at high densities, whereas both join in the opposite, low density limit. 
Note that the pion mass potential term does not directly contribute to the value of $\Lambda$, but it does so indirectly because it modifies the classical solution.

The kinetic term in the Lagrangian of an isospinning cubic crystal with a number $N_{\text{cells}} \equiv N$ of unit cells can thus be written in terms of the isospin moment of inertia $\Lambda$ \eqref{iso_moment} as
\begin{equation}
    T=\frac{1}{2}\omega_i\Lambda_{ij}\omega_j=N\frac{1}{2}\Lambda\omega_a\omega^a,
\end{equation}
and, by defining the corresponding canonical momentum $J_a=\partial L/\partial \omega^a=N\Lambda\omega_a$, we may write it in Hamiltonian form,
\begin{equation}
    H=\frac{1}{2N\Lambda}J_aJ^a.
\end{equation}
Now, following the standard canonical quantization procedure, we promote the isospin angular momentum variables to operators, so that we may diagonalise the Hamiltonian in a basis of eigenstates with a definite value of the total isospin angular momentum,
\begin{equation}
    H=\frac{\hbar^2}{2N\Lambda}J^{\rm{tot}}(J^{\rm{tot}}+1)
    \label{Hamiltonian}
\end{equation}
The total isospin angular momentum of the full crystal will be given by the product of the total number of unit cells times the total isospin of each unit cell, which can be obtained by composing the isospin of each of the cells. In the charge neutral case, all cells will have the highest possible value of isospin angular momentum, so that in each unit cell with baryon number  $B_{\text{cell}}$, the total isospin will be $\frac{1}{2}B_{\text{cell}}$, and hence the total isospin of the full crystal will be $J^{\rm{tot}}=\frac{1}{2}N B_{\text{cell}}$.

Therefore, the quantum correction to the energy (per unit cell) due to the isospin degrees of freedom will be given by (assuming $N\rightarrow \infty$)

\begin{equation}
    E_{\rm{iso}}=\frac{\hbar^2}{8\Lambda}B_{\text{cell}}^2.
    \label{Eiso}
\end{equation}
For example, for the unit cell of the FCC half-Skyrmion lattice, the isospin contribution to the energy per baryon is
\begin{equation}
    \frac{E_{\rm{iso}}}{B_{\text{cell}}}\Bigg\rvert_{\rm{FCC}}=\frac{\hbar^2}{2\Lambda}.
\end{equation}

The value of $\hbar$ in Skyrme units is important to calculate the contribution of the isospin energy, however for our choice of the parameters $\hbar = e^2/(3\pi^2) \approx 1$.

The classical Skyrmion crystal configurations can be understood as models for isospin-symmetric nuclear matter, i.e. nuclear matter with zero total isospin. Indeed, since the Skyrme Lagrangian is symmetric under isospin rotations of the chiral fields, in principle there is no distinction between nucleons in classical configurations. However, as we have seen, the quantum isospin correction to the crystal energy per baryon does depend on the difference between protons and neutrons through the total isospin number per unit cell. Hence, by considering the effect of iso-rotations over classical solutions we are effectively breaking the isospin symmetry of the static energy functional by adding a correction of quantum origin that explicitly breaks it. 
Moreover, such correction could also have been obtained through the inclusion of an isospin chemical potential. Indeed, 
we may introduce a nonzero isospin chemical potential $\mu_I$ in any chiral effective theory (and the Skyrme model in particular) in terms of a covariant derivative of the chiral fields of the form \cite{Son:2000xc}
\begin{equation}
    \partial_\mu U\rightarrow D_\mu U=\partial_\mu U-\frac{i\mu_I}{2}\delta_{\mu 0}[\tau^3,U],
\label{Covreplacement}
\end{equation}
so that, if $U$ is a static configuration, the time component of the Maurer-Cartan form becomes
\begin{equation}
    L_0=-\frac{i}{2}\mu_IU^\dagger[\tau^3,U]=-\mu_IT_3.
\end{equation}
Comparing with \cref{MCiso}, one sees that this expression is equivalent to that of an iso-rotating field with angular velocity $\omega_a=-\mu_I\delta_{3a}$. Thus, it is straightforward to obtain the isospin chemical potential for the Skyrmion crystal using its thermodynamical definition
$
    \mu_I=-\pdv{E}{n_I}
$, 
where $n_I$ is the (third component of) the isospin number density. Given that $(J^{\rm{tot}})^2=J_1^2+J_2^2+J_3^2$ and $n_I=J_3/N$, we may write the isospin energy per unit cell as
\begin{equation}
    E_{\rm{iso}}=\frac{\hbar^2 }{2\Lambda}\qty(n_I^2+\frac{J_2^2}{N^2}+\frac{J_1^2}{N^2})
\end{equation}
and then
\begin{equation}
    \mu_I=-\pdv{E_{\rm{iso}}}{n_I}=-\frac{\hbar^2 }{\Lambda}n_I.
\end{equation}

As we have seen in the previous subsection, the ground state of the Skyrme crystal is forced by the charge neutrality condition to have $(J^{\rm{tot}})^2=J_3^2$, and for a unit cell of $B_{\text{cell}}$ baryon number, the chemical potential is simply $\mu_I(L)=-\hbar^2 B_{\text{cell}}/(2\Lambda)$. This indeed coincides with the expression of $\omega_a $ in terms of the isospin angular momentum and the isospin moment of inertia. 
As a final comment, we remark that the inclusion of isospin may have non-trivial effects on the geometry of classical solitonic solutions. In particular, it was shown in \cite{Loewe:2005kq} that the isospin chemical potential may alter the stability of classical Skyrmion configurations. Hence, a thorough analysis of dense Skyrmion matter at finite isospin chemical potential should focus on its effects on the classical crystal configuration. However, as the contribution to the isospin energy per unit cell $E_{\rm{iso}}(L)$ is never dominating for any $L$, we may neglect the backreaction of the isospin term into the background and consider it simply as a (quantum) correction to the energy. 

\section{Quantum skyrmion crystals and the symmetry energy of $npe(\mu)$ matter}
We remark that the charge neutrality constraint directly fixes the total isospin number density per unit cell of the lattice to be $i=-2$,  which describes a pure neutron crystal. This corresponds to a highly isospin asymmetric state of matter, similar to what is expected to occur in the core of neutron stars. However, realistic models of the nuclear matter inside neutron stars do not predict a totally asymmetric state, but a small fraction of protons over total nucleons is generally allowed. 

In the case of nuclear matter as described by a Skyrmion crystal, the fundamental degrees of freedom are not protons and neutrons, and in fact the number of protons and/or neutrons per unit cell is ill-defined in some crystalline phases in which the baryon charge is fractionalized due to the presence of half-Skyrmions. However, we can still define the asymmetry parameter and thus the symmetry energy of a Skyrme crystal \cite{Rho:2021hrw}, which allows us to calculate the density dependence of $S$, and obtain the above defined observables \eqref{eq_symetobs}, within the Skyrme model.

\subsection{The electric charge density of pure Skyrmion matter}
The Gell-Mann-Nishijima formula tells us how to obtain the charge density of a Skyrmion field configuration,
\begin{equation}
    Z = \frac{1}{2}B + I_3 \hspace{2mm} \longrightarrow \hspace{2mm} \rho = \frac{1}{2}B^0 + \ev{I^0_3},
    \label{charge_dens}
\end{equation}
where $I^0_3$ is the time-like component of the third isospin Noether current, and the brackets represent the expectation value on the crystal quantum state. An explicit expression for (the classical version of) this current can be obtained from Noether's theorem, given that the infinitesimal version of \eqref{IsorotF} yields the transformation 
\begin{equation}
    U\rightarrow U'=U+\epsilon^k\delta U_k,\qquad \delta U_k=\frac{i}{2}[\tau^k,U],
\end{equation}
and from the definition of the corresponding Noether current,
\begin{equation}
\begin{split}
    I^{\mu}_k =\frac{i}{24\pi^2} &\left(2a \Tr\left\{ L^{\mu} T_k\right\} - 4b \Tr\left\{\left[ L^{\nu}, L_\mu\right]\left[ L^{\nu},T_k \right]\right\} - \right.\\
    &\left.-c \epsilon^{\mu\nu\alpha\beta}B_{\nu}\Tr\left\{L_{\alpha}L_{\beta}T_k\right\} \right) ,
    \label{NoetherCurr}
\end{split}
\end{equation}
where $T_k$ is the $\mathfrak{su}(2)$ current defined in \eqref{UdcommtU}. In particular, for the third isospin charge we have
\begin{equation}
    I_k^0=-R_{kj}\lambda_{ij}\omega_i ,
\end{equation}
where $R_{ab}$ is the $SO(3)$ matrix representation of the $SU(2)$ element $g$, $R_{ab}(g)=\Tr{\tau_a g^\dagger\tau_bg}$; $\lambda_{ij}$ is the inertia tensor density, defined as the integrand of \eqref{Inertia_Tensor},  and we have used the transformation property of the Pauli matrices $g^\dagger \tau_a g=R_{ab}\tau_b$.

The quantum version of $I_3^0$ can be obtained by substituting the classical variables $g$ and $\omega_i$ with the corresponding quantum operators and Weyl ordering the products of two or more non-commuting operators. In the case of the angular velocity, we will use the quantum (body-fixed) isospin angular momentum operator, $K_i=\omega_i/(\Lambda N)$. Then, the expectation value of $I_3^0$ in the isospin state $\ket{\psi}$ is given by
\begin{equation}
    \ev{I_3^0}=-\frac{\lambda_{ij}}{2\Lambda N}\bra{\psi}(K_iR_{3j}+R_{3j}K_i)\ket{\psi},
\end{equation}
where $\lambda_{ij}$ is the isospin inertia tensor density, i.e the integrand in \eqref{Inertia_Tensor}.

Furthermore, we may write $R_{3j}K_i=K_iR_{3j}-[K_i,R_{3j}]=K_iR_{3j}-i\epsilon_{ijk}R_{3k}$, and since the isospin inertia tensor density is symmetric we just have
\begin{equation}
    \ev{I_3^0}=-\frac{\lambda_{ij}}{\Lambda N}\bra{\psi}K_iR_{3j}\ket{\psi}=-\frac{\lambda_{ij}}{\Lambda N}\bra{\psi}R_{3j}K_i\ket{\psi}.
    \label{quant_iso}
\end{equation}

\begin{figure*}
    \centering
    \hspace{-0.5cm}
    \includegraphics[scale=0.35]{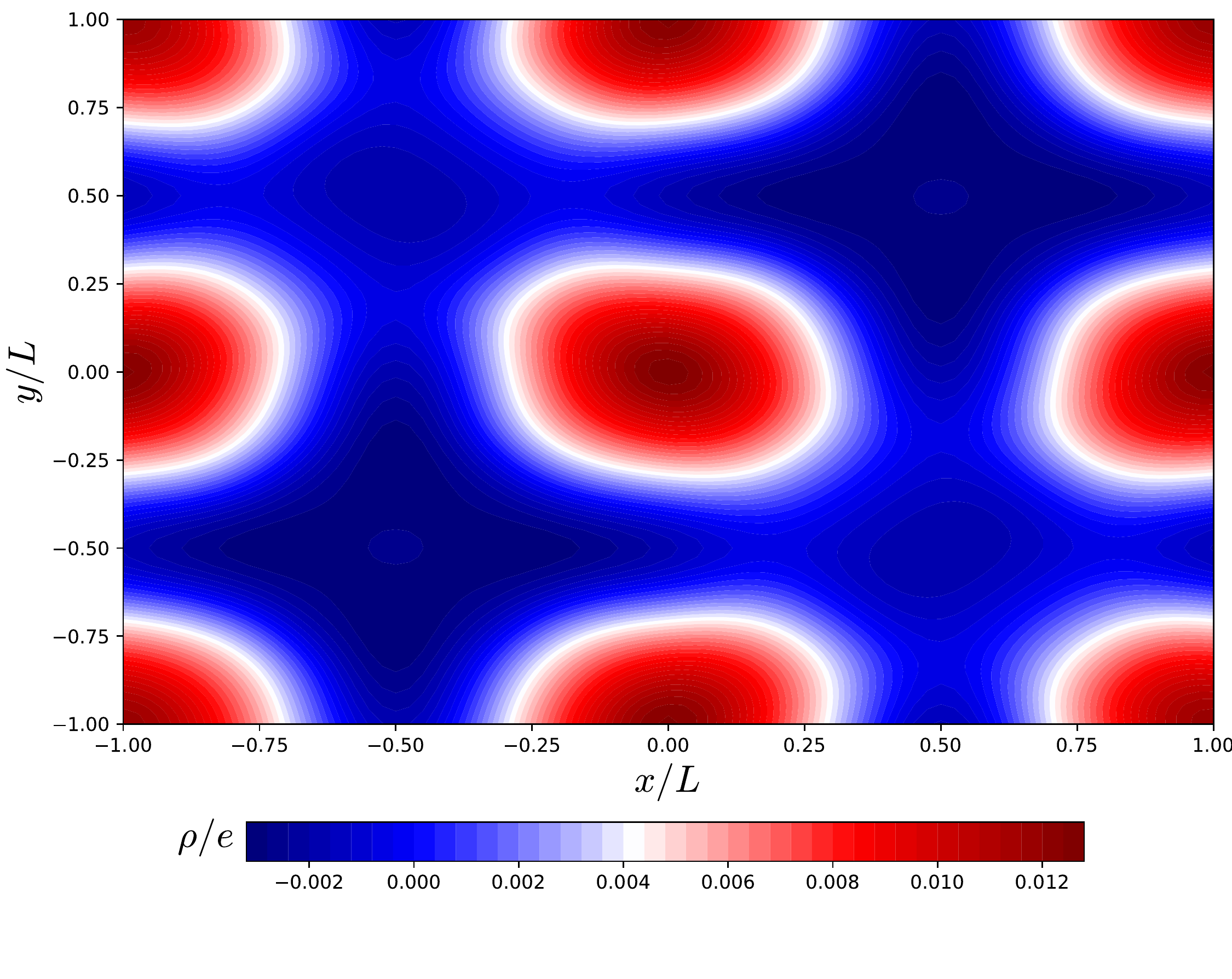}
    \caption{Electric charge density of half-Skyrmion matter in the quantum ground state.}
    \label{fig:chargedens}
\end{figure*}

As shown in the previous section, there is only one quantum state that preserves the symmetry of the unit cell for each value of the isospin quantum number, that we have denoted by $\ket{\psi^i}$,  $(i=0,1,2)$ in \eqref{quantumstates}. Therefore, for a single unit cell, the electric charge density can have three different profiles, depending on the quantum state of the cell. We are, however, interested in the case of (infinitely) many crystal cells. To obtain the charge density of the crystal, we would need to know the full quantum state. For pure skyrmion matter, we have seen that charge neutrality implies that the lowest energy state must be the tensor product $\ket{\Psi} = \bigotimes\limits_{\rm cells}\ket{\psi ^2,i_3=-2} $. Then the charge density becomes simpler, since it is just the same for each unit cell and given by \eqref{charge_dens}, where $\ev{I_0^3}$ is just the quantum isospin density for each cell, as given explicitly by \eqref{quant_iso}. Performing the calculation of the matrix element as in \cref{append_matrixelements}, we may obtain the charge density of the skyrmion crystal in its ground state, which is represented in \cref{fig:chargedens}. The figure reveals a nontrivial charge fractionalization in the half-skyrmion phase, in which half-skyrmions are positively charged while the space between them presents negative charge to keep the charge neutrality of the unit cell. 
The fact that the charge density presents a non trivial multipolar structure means that a residual Coulomb interaction between unit cells may exist. However, we have checked numerically that such a contribution is several orders of magnitude lower than the isospin correction, therefore unimportant for the total binding energy of the crystal.

\subsection{The symmetry energy}

\comment{
\subsection{Coulomb energy}
In addition of the strong nuclear interaction between nucleons, the Coulomb interaction plays an essential role in finite nuclei. Indeed, while the former keeps the system bounded, the latter produces a repulsive force that increases the binding energy. In the context of nuclear matter in neutron stars, it has been established that the competing contribution of surface terms and Coulomb interactions to the energy per baryon may result in exotic distributions of nuclei known as \emph{nuclear pasta} at densities $\rho \sim 1.4\times10^{14}{\rm g/cm^{3}}$ in the inner crust of neutron stars. 

On the other hand, as shown in the previous sections, it is precisely at such pressure at which the Skyrmion crystal reaches its lower value of energy per baryon, and starts to desintegrate into the so-called inhomogeneous phase.

Hence we will explain now how to include the Coulomb energy in the Skyrme crystal.

The Coulomb interaction in the non-relativistic case is given by:
\begin{equation}
    E_C = \frac{1}{2\epsilon_0}\int d^3x \: d^3x' \frac{\rho(x)\rho(x')}{4\pi |x - x'|},
\end{equation}
where $\epsilon_0 = \frac{1}{q_e}8.8542\: 10^{-21} \frac{1}{\text{MeV fm}}$ is the permittivity of vacuum, $q_e$ is the unit electric charge and $\rho$ represents the electromagnetic charge density. The value of $\epsilon_0$ will also change in our units in a similar way than $\hbar$:
\begin{equation}
    \epsilon_0 = \frac{8.8542\: 10^{-21}}{q_e} 3\pi^2 \frac{197.3269804}{e^2}.
\end{equation}
 
}
Let us consider a finite Skyrme crystal of $N$ unit cells, and let $B=N\times B_{\text{cell}}$, where $B_{\text{cell}}$ is the baryon number of a unit cell. We do not enforce charge neutrality at this step, and further leave unknown the quantum state $\ket{\Psi}$ of the crystal. We have seen that the total charge of this system is
    \begin{equation}
        Q=\ev{e\int d^3x \{B^0/2+I^0_3\}}=eN\qty[\frac{B_{\text{cell}}}{2}+\frac{\ev{\int I^0_3 d^3x}}{N}]
    \end{equation}
    
As argued at the end of \cref{sec:Quantization}, the total third component of isospin is a good quantum number for the total quantum state of the crystal, although this is not true for individual unit cells. In other words, the expectation value
\begin{equation}
   \ev{ I_3}=\bra{\Psi}\int I^0_3 d^3x\ket{\Psi}
\end{equation}
is well defined in an arbitrary quantum state, but $\int \ev{I^0_3} d^3x $ is not. Since we are seeking for a definition of the isospin density in the quantum theory, we may perform a mean field approximation and consider that the isospin density in an arbitrary skyrmion crystal quantum state is approximately uniform so that
\begin{equation}
    \ev{I^0_3}=\frac{\ev{I_3}}{\int d^3x}=\frac{\ev{I_3}}{NV_{\rm cell }}\doteq \frac{n_I}{V_{\rm cell}}
\end{equation}
where $n_I$ is the effective isospin charge  per unit cell in this arbitrary quantum state. We may further  consider the effective proton fraction that would yield such an isospin charge per unit cell with baryon number $B_{\text{cell}}$ to write
    \begin{equation}
       n_I=-\frac{1}{2}(1-2\gamma)B_{\text{cell}}=-\frac{B_{\text{cell}}}{2}\delta .
    \end{equation}
Hence, we may write the isospin energy per unit cell of the Skyrmion crystal in such a state in terms of the asymmetry parameter
\begin{equation}
    E_{\rm iso}=\frac{\hbar^2 B_{\text{cell}}^2}{8\Lambda}\delta^2,
    \label{Eiso}
\end{equation}
and thus the symmetry energy for Skyrme crystals is given by
\begin{equation}
    S_N(n_B)=\frac{\hbar^2 L^3}{\Lambda}n_B.
\end{equation}

\begin{figure*}
    \centering
    \includegraphics[scale=0.5]{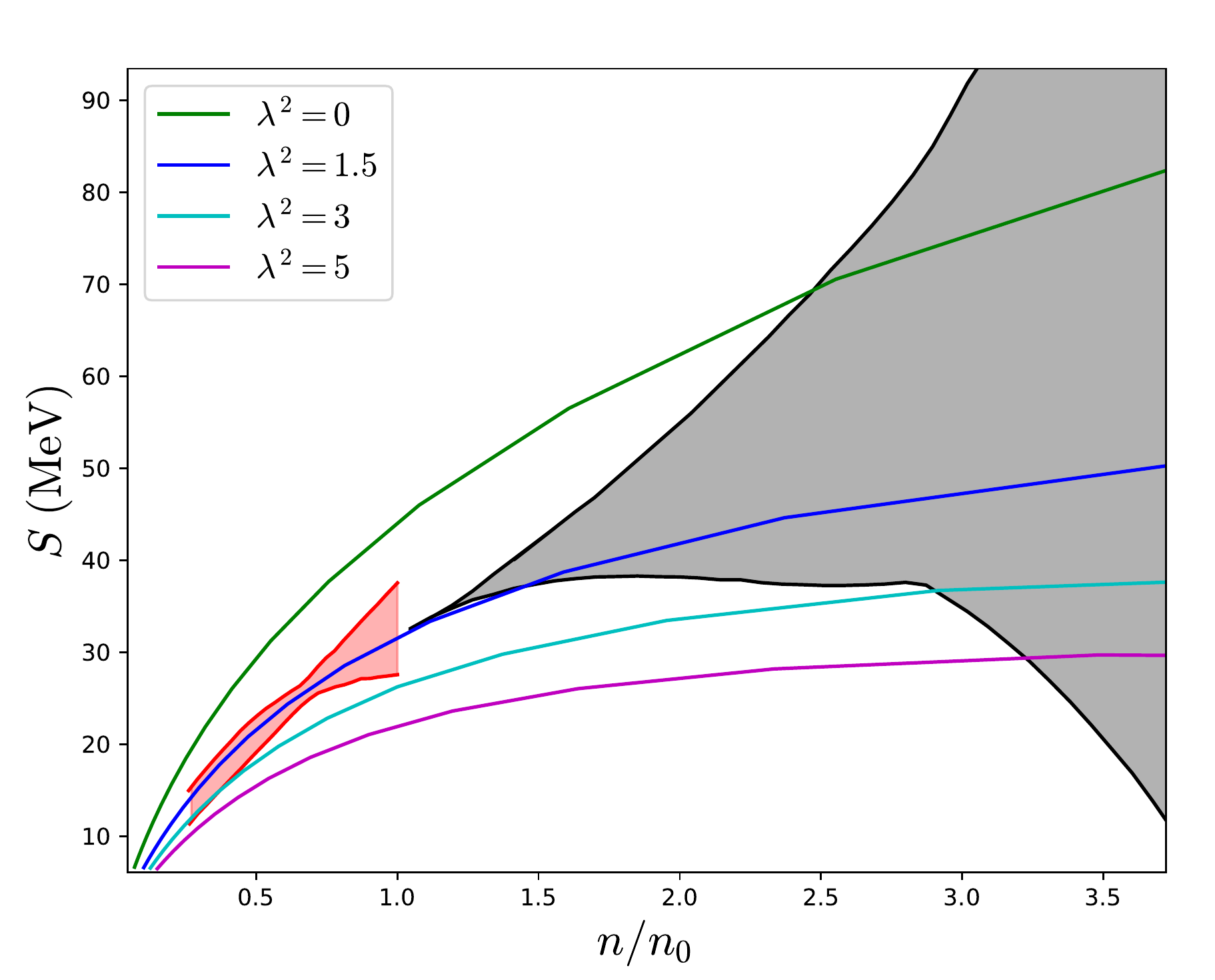}
    \caption{Symmetry energy of Skyrme crystals as a function of the density for different values of $\lambda$. Constraints of the symmetry energy at sub-saturation densities from isobaric analog states \cite{Danielewicz:2013upa} are plotted in red. The grey region corresponds to recent constraints from the analysis of neutron star observations \cite{Li:2021thg}}
    \label{fig:Symmetry_E}
\end{figure*}

In \cref{fig:Symmetry_E} we plot the symmetry energy of Skyrmion crystals for the generalized Skyrme model with different values of the sextic coupling constant. 

We can observe that, for this choice of parameters, the symmetry energy of the crystal in the $\mathcal{L}_{240}$ submodel ($\lambda^2 =0$) comes out too big with respect to the constraints at saturation density $n_0$, while the inclusion the sextic term reduces the value. We found that a value of $\lambda^2 \sim 1.5 \,{\rm MeV\,fm^3}$ fits to the correct value at saturation. Moreover, we may also compute the slope and curvature parameters at saturation for all cases, shown in \cref{Table.Bounds}. Interestingly, for values of $\lambda^2$ between $1.5 - 3$ MeV fm$^{-3}$ we find that both the saturation density and the symmetry energy parameters agree quite well with the most up-to-date experimental values. Here we have defined the saturation density as the density where the energy per baryon of the skyrmion crystal takes its minimum value.
\begin{table*}[h!]
	\centering
		\begin{tabular}{|c|c|c|c|c|}
			\hline
			$\lambda^2$ (MeV fm$^3$) & $n_0$ (fm$^{-3}$) & $S_0$ (MeV) & $L$ (MeV) & $K_\text{sym}$ (MeV) \\ \hline
			0 & 0.33 & 44.4 & 72.9 & -143 \\ \hline
			1.5 & 0.22 & 31.9 & 46.4 & -130 \\ \hline
			3 & 0.18 & 26.4 & 35.4 & -118 \\ \hline
			5 & 0.15 & 22.2 & 27.5 & -105 \\ \hline
			Exp. & 0.16 $\pm$ 0.01 & 31.7 $\pm$ 3.2 & 57.7 $\pm$ 19 & -107$\pm$ 88 \\ \hline
		\end{tabular}
		\caption{Symmetry energy coefficients in the Skyrme model with different sextic couplings. In the last row, we show the most up-to-date fiducial values of $n_0, S_0, L$ and $K_\text{sym}$ \cite{Li:2021thg}}
		\label{Table.Bounds}
\end{table*}
Hence, the constraints on the symmetry energy yield rather stringent bounds on the value of $\lambda^2$, where the precise numerical values of these bounds will, of course, depend on the choices made for the other model parameters $f_\pi$ and $e$. We remark that a lower bound for this constant can also be obtained from the maximum mass requirement of neutron star EoS \cite{Adam:2020aza}.\\

We end our discussion on the Symmetry energy of Skyrme crystals by pointing out the absence of the \emph{cusp structure} predicted by Lee et al. \cite{Lee:2010sw} in our results of $S$. Their argument (recently reviewed in \cite{Rho:2021hrw}) for the appearance of a minimum in the symmetry energy at a given density $n_{1/2}>n_0$ is based on the Skyrmion to half-Skyrmion transition, which is proposed to happen at $n_{1/2}\sim 2-3n_0$. In our pure Skyrme model setting, however, such a transition has been shown to occur in a thermodynamically unstable branch of the Skyrmion crystal phase diagram once the pion mass is taken into account  (see, eg. \cite{Park:2009bb,Adam:2021gbm}). Therefore, we do not find such a transition, as we consider a crystal of (nearly) half-Skyrmions to be the correct ground state for densities $n>n_0$. The ground state of the Skyrme model for densities $n\leq n_0$ is still not well understood, and some inhomogeneous configurations have been proposed \cite{PARK2019231,Adam:2021gbm} that point towards a complicated phase structure predicted by the Skyrme model near saturation. In particular, the transition from regular nuclear matter to a crystal of half-Skyrmions should take place in such a range of densities.

\subsection{Particle fractions of $npe\mu$ matter in $\beta$-equilibrium}
\label{sec:Protfrac}

As previously argued, any quantum state that deviates from the ground state in pure skyrme matter would lead to a divergence in the Coulomb energy in the infinite crystal limit. Indeed, an isolated system of positively charged matter is unstable due to Coulomb repulsion. Therefore, it is assumed that there exists a neutralizing background of negatively charged leptons (electrons and possibly muons), such that this Coulomb repulsion is compensated. Such a system of nuclear matter plus leptons is characterized in the equilibrium phase by two equilibrium conditions, namely the \emph{charge neutrality condition}
\begin{equation}
    n_p=\frac{Z}{V}=n_e+n_\mu ,
\label{chargecond}
\end{equation}
i.e., the densities of positively charged nucleons (protons) and negatively charged leptons (electrons and muons) are equal, and the \emph{$\beta$-equilibrium condition}
\begin{equation}
    \mu_n=\mu_p+\mu_l\implies\mu_I=\mu_l,\qquad l=e,\mu ,
\label{betacond}
\end{equation}
i.e., the isospin chemical potential must equal that of charged leptons, such that the neutron decay and electron capture processes
\begin{equation}
    n\rightarrow p+l+\bar{\nu}_l\quad ,\quad p+l\rightarrow n+\nu_l
\end{equation}
take place at the same rate. Moreover, leptons inside a neutron star are usually described as a non-interacting, highly degenerate fermi gas, so that the chemical potential for each type of leptons can be written
\begin{equation}
    \mu_l=\sqrt{(\hbar k_{F})^2+m_l^2}
\end{equation}
where $k_{F}=(3\pi^2n_l)^{1/3}$ is the corresponding Fermi momentum, and $m_l$ is the mass of the corresponding lepton. Indeed, for sufficiently large densities the electron chemical potential will be larger than the muon mass, $\mu_e\geq m_\mu$, and the appearance of muons in the system will be energetically favorable. We may now estimate the total proton fraction by enforcing both charge neutrality and beta equilibrium. Let us start by neglecting the contribution of muons to the charge density. Then, from the charge neutrality condition \eqref{chargecond}, we relate the electron density to the proton fraction parameter, $n_e=\gamma B_{\text{cell}} /(2L)^3$, and the $\beta$ equilibrium condition yields an equation that implicitly defines $\gamma$ as a function of the lattice length parameter,
\begin{equation}
    \frac{\hbar L}{\Lambda}(1-2\gamma)=\qty(\frac{3\pi^2}{B_{\text{cell}}^2})^{1/3}\gamma^{1/3}
\end{equation}
where we have also made the ultrarelativistic electron approximation, i.e. $m_l/k_F\simeq 0$. 


Including the muon contribution to the charge density yields a slightly more complicated expression for the $\beta$-equilibrium condition, given by
\begin{equation}
    \frac{\hbar B_{\text{cell}}}{2\Lambda}(1-2\gamma)=\qty[3\pi^2\qty(\frac{\gamma B_{\text{cell}}}{8L^3}-n_\mu)]^{\tfrac{1}{3}},
\end{equation}
where 
\begin{equation}
    n_\mu=\frac{1}{3\pi^2}\qty[\qty(\frac{\hbar B_{\text{cell}}(1-2\gamma)}{2\Lambda})^2-\left(\frac{m_\mu}{\hbar}\right)^2]^{\tfrac{3}{2} }.
\end{equation}
 On the other hand, the proton fraction inside the beta-equilibrated matter also determines whether a proto-neutron star will go through a cooling epoch via neutrino emission through the direct Urca (DU) process $n\rightarrow p +e+\bar \nu_e$, which is expected to occur if the proton fraction reaches a critical value, $\gamma_p> x_{DU}$, the so-called DU-threshold \cite{PhysRevLett.66.2701,Klahn:2006ir}. As the DU process allows for an enhanced cooling rate of NS, whether it takes place or not in the hot core of proto-neutron stars or during the merge of binary NS systems \cite{universe7110399} would determine the proton fraction (hence, the symmetry energy) of matter at ultra-high densities. However, it is not clear whether such enhanced cooling actually takes place, although there is recent evidence that supports it \cite{Brown:2017gxd}.

 In $npe\mu$ matter, the DU threshold is given by \cite{Klahn:2006ir}
 
 \begin{equation}
     x_{DU}=\frac{1}{1+(1+(\frac{n_e}{n_e+n_\mu})^{1/3})^3}.
 \end{equation}

\comment{
Once the proton fraction and the asymmetry parameter are obtained, the total energy of the system may be also calculated as
\begin{equation}
    E= E_{\rm class} + E_{\rm iso}(\delta)+E_{e}(\gamma)+E_{\mu}(\gamma)
\end{equation}
where $E_{\rm clas}$ is the classical energy of the Skyrme field, $E_{\rm iso}$ is given by \cref{Eiso} and $E_{\rm lep}$ is the energy density of a relativistic lepton gas with mass $m_l$ at zero temperature 
\begin{align}
    E_{\rm lep}&=\int_0^{k_f}\frac{k^2dk}{\pi^2}\sqrt{k^2+m_{l}^2}=\\
    &=\frac{m^4_l}{8\pi^2}\qty[x_r(1+2x_r^2)\sqrt{1+x_r^2}-\ln{x_r+\sqrt{1+x_r^2}}],\notag
\end{align}
where $x_r=k_F/m_l$.
For densities $n\geq n_0$, the electrons become ultra-relativistic, i.e. $m_e\ll k_{Fe}$  and the corresponding energy becomes
\begin{equation}
    E_e\simeq \frac{3}{4}k_{Fe}n_e.
\end{equation}
}

\begin{figure*}[htb!]

    \centering
    \includegraphics[scale=0.45]{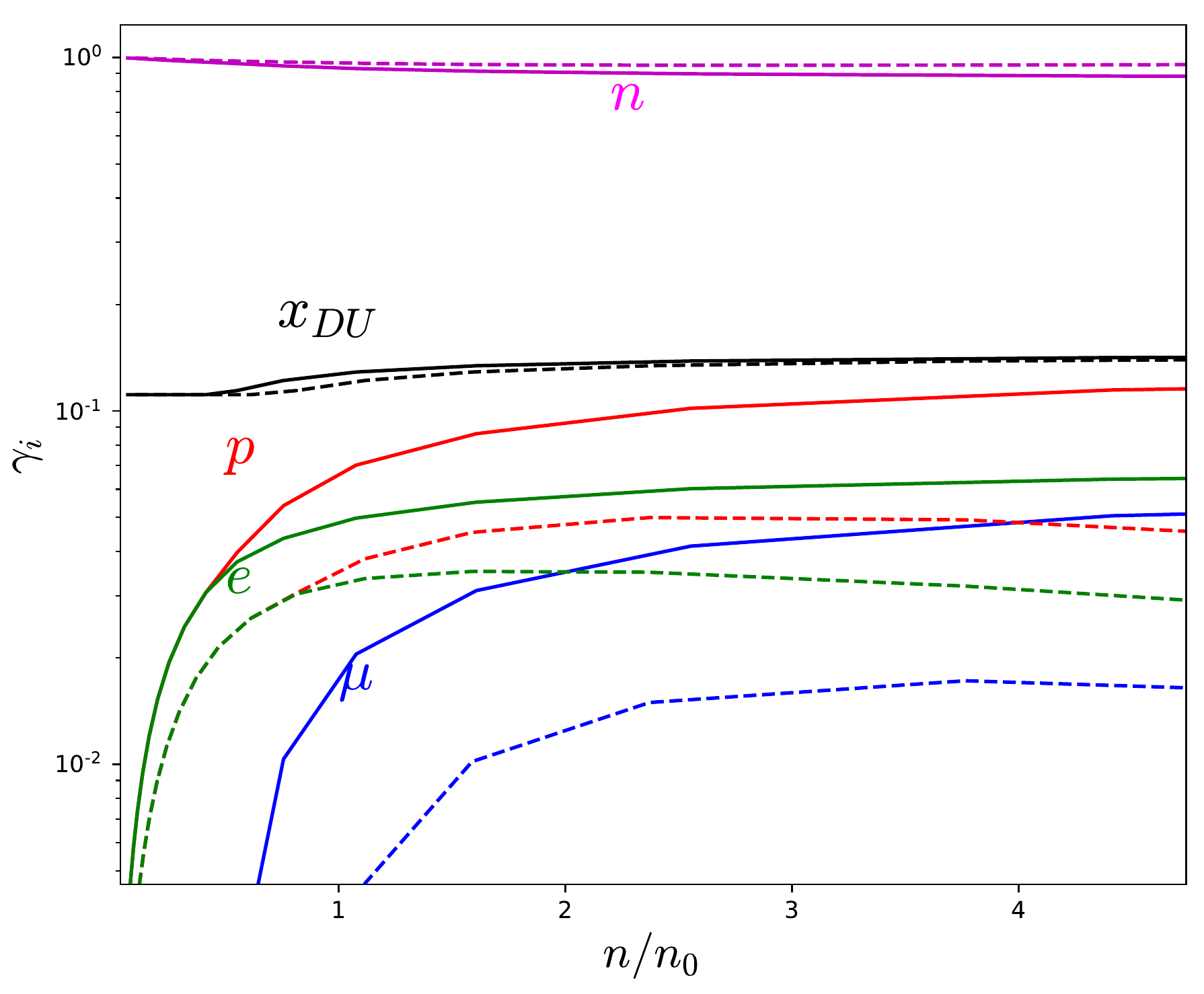}
    \caption{Fraction density $\gamma_i$ for each particle as a function of the baryon density for $\lambda^2=0$ (solid) and $\lambda^2=1.5$ (dashed). The corresponding DU threshold is also shown in black.}
    \label{fig:Protonfractionse}
    
\end{figure*}
The particle populations $\gamma_i$ in the beta-equilibrated Skyrmion matter are shown in \cref{fig:Protonfractionse} for the cases $\lambda^2=0,1.5$ MeV fm$^3$. 

In both cases, a persistent population of protons and leptons with increasing nucleon density is expected,  although in the case with sextic term we see that the fraction of charged particles is smaller. This is the impact of the sextic term, since it is much easier to convert protons into neutrons due to the lower symmetry energy. Finally, the DU-threshold is not reached in any case for the values of the parameters ($f_{\pi}, e, \lambda^2$) that we have chosen. However, one should not take this fact as a prediction of the Skyrme model, as it strongly depends on the parameter values. Also, it is generally assumed that around $2-3$ times the nuclear saturation density, additional degrees of freedom (strange baryons) become important for the description of nuclear matter, which in particular may affect the proton fraction at such densities. 

\section{Conclusions}
In this paper, we have applied standard quantization methods to find the quantum ground state of classical skyrmion crystal configurations. The quantization of the isospin degrees of freedom has allowed us to determine how the isospin contribution to the energy (hence the isospin chemical potential) depends on the crystal density. Then, we have shown that the effects of a finite isospin asymmetry can be included once the divergence in Coulomb energy of the infinite crystal is avoided by the inclusion of a neutralising leptonic background, yielding a particularly simple model for studying the high density regime of isospin asymmetric nuclear matter. 


In conclusion, the Skyrme model yields a concrete, nonperturbative framework in which not only the energy of symmetric matter can be calculated at high densities, but also the corresponding symmetry energy can be computed. This extends the list of physical, dense nuclear matter observables which can be described by the model. Further, we can fit the parameters of the model to physical observables like the symmetry energy at saturation, in order to be able to make predictions about the nuclear matter EoS at higher densities within the Skyrme model. 
Concretely, we found that a simple scan of values for the sextic term coupling constant allows us to fit the symmetry energy well within the current uncertainties. However, the saturation energy still comes out a bit smaller than its experimental value. 
This is a universal feature of the Skyrme model, which usually yields too small values for the energy per baryon of nuclei and nuclear matter, i.e., a too strongly bound nuclear matter.
Our choice for these parameter values is mainly motivated by the possibility to compare with other Skyrme model calculations, being our main objective to provide a clear discussion of the most important physical properties of quantized skyrmion crystals, including the contribution from the sextic term. The determination of the best set of parameter values to describe nuclear matter and NS should include the calculation of NS properties and a careful comparison with the most recent observational data like the ones inferred from the LIGO and NICER detections, as was done, e.g., in \cite{Adam:2020yfv}, \cite{Adam:2020aza}, and is out of the scope of this work.

Finally, the results of this paper open the door for a further improvement in the description of the equation of state for dense Skyrmion matter, including the contribution of strange degrees of freedom, such as kaon condensates or hyperonic matter which can be done systematically in the Skyrme model \cite{callan1985bound,Blom1989HyperonsAB}. Indeed, hyperons and strange mesons are believed to constitute a non-negligible fraction of the total particle content of ultra-dense matter, therefore they may be important for an accurate determination of the proton fraction at densities $n\gtrsim 2-3n_0$ \cite{PhysRevC.81.035203}.

\begin{acknowledgements}
The authors would like to thank C. Naya and N. Manton for helpful discussions and comments.

Further, the authors acknowledge financial support from the Ministry of Education, Culture, and Sports, Spain (Grant No. PID2020-119632GB-I00), the Xunta de Galicia (Grant No. INCITE09.296.035PR and Centro singular de investigación de Galicia accreditation 2019-2022), the Spanish Consolider-Ingenio 2010 Programme CPAN (CSD2007-00042), and the European Union ERDF.
AW is supported by the Polish National Science Centre,
grant NCN 2020/39/B/ST2/01553.
AGMC is grateful to the Spanish Ministry of Science, Innovation and Universities, and the European Social Fund for the funding of his predoctoral research activity (\emph{Ayuda para contratos predoctorales para la formaci\'on de doctores} 2019). MHG is also grateful to the Xunta de Galicia (Consellería de Cultura, Educación y Universidad) for the funding of his predoctoral activity through \emph{Programa de ayudas a la etapa predoctoral} 2021.
\end{acknowledgements}

\appendix
\section{Calculation of matrix elements}
\label{append_matrixelements}
The physical Hilbert space is spanned by the basis $\ket{X}=\ket{ii_3k_3}$ of states with fixed total isospin and third components of body-fixed and space-fixed isospin. We may parametrize the isospin collective coordinates as an element $A\in SU(2)$ by $A=a_0+ia_k\tau_k$, where $a=(a_0,\Vec{a})$ is subject to the constraint $a_0^2+a_ka_k=1$. Thus, the coordinate vector $a$ can be thought of parametrizing a four-dimensional sphere, and hence, in the $\ket{A}$ representation, the wave function of each of the basis states corresponds to the (hyper-)spherical harmonics on $S^3$:
\begin{equation}
    \psi(A)=\braket{A}{ii_3k_3}=Y_{i_3,k_3}^{2i}(a),
\end{equation}
with
\begin{equation}
    Y_{m,m'}^{2j}(A)=[(2j+1)/8\pi^2]^{1/2}{D}^j_{m,m'}(A),
    \label{hyperspherics}
\end{equation}
where ${D}^j_{m,m'}(A)$ are the Wigner's $D$-matrices, associated to an irreducible, spin-$j$ representation of the $SU(2)$ group. Let us write its explicit expression in terms of the matrix elements of an arbitrary $2\times 2$ matrix $B$ \cite{Bander:1965rz}:
\begin{align}
    &\hspace{1cm}B=\mqty(a&b\\c&d)\implies \mathscr{D}^j_{m,m'}(B),\qquad {\rm with}\notag\\
    &\mathscr{D}^j_{m,m'}(B)=[(j+m)!(j-m)!(j+m')!(j-m)!]^{1/2}\times\notag\\
    &\hspace{1cm}\times\sum\limits_{n_i>0}\frac{a^{n_1}b^{n_2}c^{n_3}d^{n_4}}{n_1!n_2!n_3!n_4!},
    \label{Wignerstates}
\end{align}
and the constants $n_i$ are related through ${n_1+n_2=j-m}$, ${n_3+n_4=j+m}$, ${n_1+n_3=j+m'}$, $n_2+n_4=j-m'$.
Indeed, the wavefunctions \eqref{hyperspherics} are correctly normalized under the corresponding inner product, 
\begin{equation}
    \int_{SU(2)} (Y_{i_3,k_3}^{2j}(A))^*Y_{i'_3,k'_3}^{2j'}(A) d\mu(A)=\delta_{jj'}\delta_{i_3i_3'}\delta_{k_3k_3'}
    \label{Innerharms}
\end{equation}
where $d\mu(A)$ is the Haar measure of $SU(2)$, which can be written in this case as a volume element in terms of three angular variables parametrizing the three-sphere $S^3$. We may choose these angles to be the standard hyperspherical coordinates for a unit (Euclidean) 4-vector $a$:
\begin{align}
    a_0=\cos\chi,\quad a_1=\sin\chi\sin\theta\cos\phi,\notag\\ a_2=\sin\chi\sin\theta\sin\phi,\quad a_3= \sin\chi\cos\theta
\end{align}
where the angles take the values $\chi,\theta\in [0,\pi]$, $\phi\in[0,2\pi]$. The Haar measure is then
\begin{equation}
    d\mu(A[\chi,\theta,\phi])=\sin^2\chi\sin\theta d\chi d\theta d\phi,
\end{equation}
and the states $D_{m,m'}^{j}(A[\chi,\theta,\phi])$ are obtained from \cref{Wignerstates} given that $a=ia_0-a_3$, $b=ia_2-a_1$, $c=-b^*$, $d=a^*$.
Once we have characterized all the basis states, we are ready to calculate  the matrix elements of any observable that can be expressed as an operator $\hat{F}(A)$ over the Hilbert space $\mathcal{H}$. Indeed, for any pair of basis sates $\ket{ii_3k_3},\ket{i'i_3'k_3'}$, we may write the matrix element as an integral over $SU(2)$:
\begin{equation}
    \hspace{-5mm}\bra{ii_3k_3}\hat{F}(A)\ket{i'i_3'k_3'}=\int_{SU(2)}(Y_{i_3,k_3}^{2i}(A))^* F(A)Y_{i'_3,k'_3}^{2i'}(A) d\mu(A).
\end{equation}
or equivalently as an integral in terms of the hyperspherical angles.

\section{Isospin inertia tensor of Skyrme crystals}
\label{append:inertia-tensor}

All crystalline configurations that we have considered in section \ref{Sec: Crystal ansatz} share the same basic cubic symmetry group, generated by the $A_1$ and $A_2$ transformations. This implies that Skyrmion crystals present a cubic symmetry in isospin space as well as in real space. Therefore, the isospin inertia tensor of such configurations becomes proportional to the identity, i.e. $\Lambda^{\rm{crystal}}_{ij}=\Lambda\delta_{ij}$. We will now prove this statement and obtain an explicit expression for the eigenvalue of the isospin inertia tensor. To do so, let us consider the three terms in the rhs of \eqref{Inertia_Tensor} separately.

\paragraph{Quadratic term.}
We need to calculate the following trace:
\begin{equation}
\begin{split}
    \Tr\{T_iT_j\}&=-2T_i^aT_j^a=
    2(\pi_i\pi_j-\pi_b\pi_b\delta_{ij})
    \label{expre_quadratic}
\end{split}
\end{equation}
This expression has the form of the standard definition of the inertia tensor. It is indeed symmetric, but not diagonal. However, we have not used the cubic symmetry of the configuration yet. Let us see that, as in the case of a cubic rigid body, the moment of inertia becomes proportional to the identity due to its symmetry. Consider the integral of $\pi_i\pi_j$ over a unit cell $\Omega$ of size $2L$:
\begin{equation}
    I_{ij}=\int_{-L}^{L}dx\int_{-L}^{L}dy\int_{-L}^{L}dz \pi_i\pi_j .
\end{equation}
It is straightforward to see that symmetry $A_1$ implies that $I_{ij}=0$ for $i\neq j$ and that symmetry $A_2$ implies $I_{xx}=I_{yy}=I_{zz}$, so that we may write
\begin{equation}
    I_{ij}=\delta_{ij}\frac{1}{3}\int_{\Omega} \pi_a\pi^a d^3x,
\end{equation}
and hence the total contribution of the quadratic term to the inertia tensor of the crystal can be written
\begin{equation}
    2a\Lambda^{(2)}_{ij}=2a\Lambda^{(2)}\delta_{ij},\quad \text{with}\quad \Lambda^{(2)}=-\frac{4}{3}\int_{\Omega}\pi_a\pi^ad^3x.
\end{equation}
\paragraph{Quartic term.}
The quartic term is proportional to:
\begin{equation}
    \Tr\{[T_i,L_k][T_j,L_k]\}=8[T^a_iT^a_jL_k^bL_k^b-T_i^aL_k^aT^b_jL_k^b]
\end{equation}
Noting that:
\begin{equation}
    \begin{split}
        T^a_iT^a_j&=\vec{\pi}^2\delta_{ij}-\pi_i\pi_j,\\
        T_i^aL_k^a&=i[2\sigma(\pi_i\vec{\pi}\partial_k\vec{\pi}-\vec{\pi}^2\partial_k\pi_i)+(\vec{\pi}\times\partial_k\vec{\pi})_i], \\
        L^a_kL_k^a&=-[(\sigma\partial_k\pi_a-\pi_a\partial_k\sigma)^2+\vec{\pi}^2(\partial_k\vec{\pi})^2-(\vec{\pi}\partial_k\vec{\pi})^2],
    \end{split}
\end{equation}
we have
\begin{widetext}
\begin{equation}
\begin{split}
    \Tr\{[T_i,L_k][T_j,L_k]\}=&-8(\vec{\pi}^2\delta_{ij}-\pi_i\pi_j)[(\sigma\partial_k\pi_a-\pi_a\partial_k\sigma)^2+\vec{\pi}^2(\partial_k\vec{\pi})^2-(\vec{\pi}\partial_k\vec{\pi})^2]+8(\vec{\pi}\times\partial_k\vec{\pi})_i(\vec{\pi}\times\partial_k\vec{\pi})_j\\
    +&16\sigma[(\vec{\pi}\times\partial_k\vec{\pi})_i(\pi_j\vec{\pi}\partial_k\vec{\pi}-\vec{\pi}^2\partial_k\pi_j)+(\vec{\pi}\times\partial_k\vec{\pi})_j(\pi_i\vec{\pi}\partial_k\vec{\pi}-\vec{\pi}^2\partial_k\pi_i)]+\\
    +&32\sigma^2[(\vec{\pi}\partial_k\vec{\pi})^2\pi_i\pi_j+\vec{\pi}^4\partial_k\pi_i\partial_k\pi_j-(\vec{\pi}\partial_k\vec{\pi})\vec{\pi}^2(\pi_i\partial_k\pi_j+\pi_j\partial_k\pi_i)]
\end{split}
\end{equation}
\end{widetext}
this expression is again symmetric in $i,j$ and can be further reduced when taking into account the $A_1$ and $A_2$ symmetries of crystal configurations. Indeed, following the same reasoning that in the previous case, one can show, by making use of such symmetries, that for any scalar function $F(\vec{x})$ and any vector function $G_k(\vec{x})$, one has
\begin{equation}
\begin{split}
    &\int_{\Omega}F\pi_i\pi_j d^3x=\delta_{ij}\frac{1}{3}\int_{\Omega}F\vec{\pi}^2d^3x, \\ &\int_{\Omega}F\partial_k\pi_i\partial_k\pi_j d^3x=\delta_{ij}\frac{1}{3}\int_{\Omega}F(\partial_k\vec{\pi})^2d^3x,\\
    &\int_{\Omega}G_k\pi_i\partial_k\pi_j d^3x=\delta_{ij}\frac{1}{3}\int_{\Omega}G_k\vec{\pi}\partial_k\vec{\pi}d^3x,\\
    &\int_{\Omega}(\vec{\pi}\times\partial_k\vec{\pi})_i(\vec{\pi}\times\partial_k\vec{\pi})_j d^3x=\delta_{ij}\frac{1}{3}\int_{\Omega}(\vec{\pi}\times\partial_k\vec{\pi})^2d^3x,\\
    &\int_{\Omega}F\partial_k\pi_i(\vec{\pi}\times\partial_k\vec{\pi})_j d^3x=\int_{\Omega}G_k\pi_i (\vec{\pi}\times\partial_k\vec{\pi})_jd^3x=0.
\end{split}
\label{relations}
\end{equation}
Therefore, the total contribution to the inertia tensor of the quadratic term is indeed proportional to the identity, and given by
\begin{equation}
    -4b\Lambda^{(4)}_{ij}=-4b\Lambda^{(4)}\delta_{ij}, \quad \rm{with}
\end{equation}
\begin{equation}
\hspace{-1cm}\begin{split}
    \Lambda^{(4)}=-&\frac{8}{3}\int_\Omega\!\!d^3x
    \left\{2\vec{\pi}^2[(\sigma\partial_k\pi_a-\pi_a\partial_k\sigma)^2+\vec{\pi}^2(\partial_k\vec{\pi})^2-(\vec{\pi}\partial_k\vec{\pi})^2]\right.\\[2mm]
    -&\left.(\vec{\pi}\times\partial_k\vec{\pi})^2-4\sigma^2[(\vec{\pi}\partial_k\vec{\pi})\vec{\pi}-\vec{\pi}^2\partial_k\vec{\pi}]^2
    \right\}.
\end{split}
\end{equation}
\paragraph{Sextic term.}
For the sextic term contribution we must calculate the following expression:
\begin{equation}
    \varepsilon^{abc}\Tr\{T_iL_bL_c\}\varepsilon_{ars}\Tr\{T_jL_rL_s\}=
    8\varepsilon_{lmn}\varepsilon_{pqr}T_i^lL_a^mL_b^nT_j^pL_a^qL_b^r.
\end{equation}
Given that
\begin{equation}
\begin{split}
    \varepsilon_{lmn}T_i^lL_a^mL_b^n=&-\Big [A_{ab}\pi_i+B_b\partial_a\pi_i-B_a\partial_b\pi_i+C_b(\vec{\pi}\times\partial_a\vec{\pi})_i-\\
    &-C_a(\vec{\pi}\times\partial_b\vec{\pi})_i+D(\partial_a\vec{\pi}\times\partial_b\vec{\pi})_i\Big],
\end{split}
\end{equation}

where
\begin{equation}
    \begin{split}
        A_{ab}&=\vec{\pi}^2(\partial_b\sigma\vec{\pi}\cdot\partial_a\vec{\pi}-\partial_a\sigma\vec{\pi}\cdot\partial_b\vec{\pi})-\sigma^2\vec{\pi}\cdot(\partial_a\vec{\pi}\times\partial_b\vec{\pi}),\\
        B_a&=\sigma\vec{\pi}^2\vec{\pi}\cdot \partial_a\vec{\pi}-\vec{\pi}^4\partial_a\sigma,\\
        C_a&=\sigma^2\vec{\pi}\cdot\partial_a\vec{\pi}-2\vec{\pi}^2\sigma\partial_a\sigma,\quad D=\sigma^2\vec{\pi}^2,
    \end{split}
\end{equation}
we are ready to write the contribution of the sextic term to the inertia tensor:

\begin{equation}
    -\frac{c}{32\pi^4}\Lambda^{(6)}_{ij}=-\frac{c}{32\pi^4}\Lambda^{(6)}\delta_{ij},
\end{equation}
with
\begin{widetext}
\begin{equation}
\begin{split}
        \Lambda^{(6)}&\!=\!\frac{1}{3}\int_\Omega\!\!d^3x\big\{A_{ab}A_{ab}\vec{\pi}^2+2A_{ab}B_{b}\vec{\pi}\partial_a\vec{\pi}-2A_{ab}B_{a}\vec{\pi}\partial_b\vec{\pi}+2B_bB_b\partial_a\vec{\pi}\partial_a\vec{\pi}-2B_aB_b\partial_b\vec{\pi}\partial_a\vec{\pi}+\\[2mm]
        &+2C_aC_a(\vec{\pi}\times\partial_b\vec{\pi})(\vec{\pi}\times\partial_b\vec{\pi})-2C_aC_b(\vec{\pi}\times\partial_a\vec{\pi})(\vec{\pi}\times\partial_b\vec{\pi})+\\[2mm]
        &+D^2(\partial_a\vec{\pi}\times\partial_b\vec{\pi})(\partial_a\vec{\pi}\times\partial_b\vec{\pi})+4C_aD(\vec{\pi}\times\partial_b\vec{\pi})(\partial_a\vec{\pi}\times\partial_b\vec{\pi}) + A_{ab}D\vec{\pi}(\partial_a\vec{\pi}\times\partial_b\vec{\pi})\Big\}.
\end{split}
\end{equation}
\end{widetext}
To obtain this expression, we have used again  \eqref{relations}.

\bibliography{biblio}
\end{document}